\documentclass[aps,nofootinbib,prd,eqsecnum,showpacs,showkeys,preprintnumbers,twocolumn]{revtex4-1}
\usepackage{textcomp}
\usepackage{amsmath}
\usepackage{amssymb}
\usepackage{esint}

\makeatletter

\usepackage{graphicx}
\usepackage{amsfonts}
\usepackage{color}
\usepackage{bm}
\usepackage{mathrsfs}
\usepackage{epstopdf}
\usepackage{url}
\usepackage{footnote}
\usepackage{dsfont}
\usepackage{ulem}

\newcommand*{\rom}[1]{\expandafter\@slowromancap\romannumeral #1@}

\newcommand{\be}{\begin{equation}}
  \newcommand{\ee}{\end{equation}}
\newcommand{\ben}{\begin{eqnarray*}}
  \newcommand{\een}{\end{eqnarray*}}
\newcommand{\bea}{\begin{eqnarray}}
  \newcommand{\eea}{\end{eqnarray}}
\newcommand{\bdm}{\begin{displaymath}}
  \newcommand{\edm}{\end{displaymath}}
\newcommand{\ba}{\begin{align}}
  \newcommand{\ea}{\end{align}}

\makeatother

\begin{document}

\title{Electrodynamics and spacetime geometry I: Foundations}

\author{Francisco Cabral}
\email{cosmocabral@gmail.com}
\affiliation{Instituto de Astrof\'{\i}sica e Ci\^{e}ncias do Espa\c{c}o, Faculdade de
Ci\^encias da Universidade de Lisboa, Edif\'{\i}cio C8, Campo Grande,
P-1749-016 Lisbon, Portugal.}

\author{Francisco S. N. Lobo}
\email{fslobo@fc.ul.pt}
\affiliation{Instituto de Astrof\'{\i}sica e Ci\^{e}ncias do Espa\c{c}o, Faculdade de
Ci\^encias da Universidade de Lisboa, Edif\'{\i}cio C8, Campo Grande,
P-1749-016 Lisbon, Portugal.}

\date{\today}

\begin{abstract}
We explore the intimate connection between spacetime geometry and electrodynamics. This link is already implicit in the constitutive relations between the field strengths and excitations, which are an essential part of the axiomatic structure of electromagnetism, clearly formulated via integration theory and differential forms. We briefly review the foundations of electromagnetism based on charge and magnetic flux conservation, the Lorentz force and the constitutive relations which introduce the spacetime metric. We then proceed with the tensor formulation by assuming local, linear, homogeneous and isotropic constitutive relations, and explore the physical, observable consequences of Maxwell's equations in curved spacetime. 
The field equations, charge conservation and the Lorentz force are explicitly expressed in general (pseudo) Riemanian manifolds. The generalized Gauss and Maxwell-Amp\`{e}re laws, as well as the wave equations, reveal potentially interesting astrophysical applications. In all cases new electromagnetic couplings and related phenomena are induced by spacetime curvature. The implications and possible applications for gravity waves detection are briefly addressed. 
At the foundational level, we discuss the possibility of generalizing the vacuum constitutive relations, by relaxing the fixed conditions of homogeneity and isotropy, and by assuming that the symmetry properties of the electro-vacuum follow the spacetime isometries. The implications of this extension are briefly discussed in the context of the intimate connection between electromagnetism and the geometry (and causal structure) of spacetime. 
 
\end{abstract}



\maketitle




\section{Introduction}

The classical field theory of electromagnetism lies at the very heart of profound developments in our understanding of physics. Before the conceptual revolutions of special relativity, general relativity and quantum theory, the seeds planted by the works of Faraday and Maxwell led to the establishment of the important concept of a physical field while preparing at the same time the conditions for the advent of both special relativity and quantum field theory. Indeed, it was the electrodynamics of moving objects that inspired Einstein's work on special relativity and it was the form of Maxwell's equations that motivated and guided Lorentz, Poincar\'{e} and Einstein to derive the (Lorentz and Poincar\'{e}) spacetime transformations. This in turn led to the revolutionary spacetime unification of Minkowski. 
In fact, Maxwell's equations were successfully incorporated within quantum electrodynamics and played a key role in the development of general Yang-Mills gauge models for the fundamental interactions in quantum field theories. It is well known that in this context, the Maxwell fields served as a prototype for understanding the deep relation between (gauge) symmetries and the dynamics of fundamental physical fields and interactions. On the other hand, the remarkable work of Noether allowed the understanding of the link between these so-called internal symmetries and conserved quantities. 

Nevertheless, it is worth recalling that the development of gauge symmetries had its routes in the influential work of Weyl (1918) on gravity, soon after the formulation of General Relativity in 1915 (see \cite{unifiedfield} for an historical review). Weyl generalized the gravitational theory by assuming that the light cones have the principal relevance, while abandoning the absoluteness of spacetime distances. Accordingly, in his theory the conformal (or causal) structure of spacetime is invariant while the metric $\bold{g}$ is only fixed up to a proportionality leading to a (gauge) freedom, $\bold{g}\rightarrow \lambda \bold{g}$. A given choice provides a certain gauge that allows spatial and time intervals to be determined. 
With this idea, Weyl was able to incorporate Maxwell's equations in the spacetime geometry by introducing an additional structure besides the conformal: the gauge connection (or bundle connection). The set of all possible Lorentzian metrics (related by a conformal gauge transformation) sharing the same local light cone provided the local fibres of a gauge bundle over the base spacetime manifold and a bundle connection was required. The electromagnetic potential played the role of this connection which was incorporated in the covariant derivative and the electromagnetic field tensor was the curvature of the gauge connection. Therefore, this was not only one of the first early serious attempts to intimately link the electromagnetic field with spacetime geometry, in search for a unified field theory \cite{unifiedfield}, but it also represented the very birth of gauge theories in the physics of interactions. Weyl's emphasis on the light cone and therefore on the casual structure of spacetime echoes in some sense in modern ideas of gauge theories of gravitation (see \citep{Blagojevic:2012bc}), since the local conformal gauge group is more general than the Lorentz, the Poincar\'{e} or even the so called Weyl group. The 15 parametric conformal group includes the Poincar\'{e} sub group plus dilatations and proper conformal transformations, where the last two break the line element invariace while preserving the light cone. The original ideas of Weyl changed, but the fundamental link between gauge symmetries and the dynamics, established through geometrical reasoning, namely, through gauge or bundle connections, remained in modern Yang-Mills theories.

The intimate link between electromagnetism and spacetime geometry, and therefore gravity, is one of the most relevant topics of classical field theories. On the one hand, since electromagnetic fields have energy-momentum they gravitate, affecting spacetime geometry. On the other hand, light rays propagate along null geodesics, which express an important link between the causal structure of spacetime and the propagation of electromagnetic fields. The notion of causality is fundamental in physics and the idea that it is profoundly associated to electrodynamics gives this classical field theory a special relevance. Such a relation seems to be unique, since photons are now viewed as the only massless particles of the standard model of elementary particles, and therefore are the only ones that can provide an experimental study of the null cones. 
Although the light cone first appeared within Minkowski spacetime, and Maxwell's equations were the first relativistic field equations, these can be shown to have a pre-metric formulation \cite{Gronwald:2005tv,Hehl:2000pe,Hehl:2005hu,Hehl:1999bt,Hehl:2008sm,Hehl:2004jn}, while the light cone can be derivable from electrodynamics \cite{Hehl:2005hu,Lammerzahl:2004ww,Itin:2004qr,Hehl:2004yk}. In fact, in the spacetime framework, Maxwell's equations developed naturally into Cartan's exterior calculus of differential forms and in this formalism the field equations are indeed fully general, coordinate-free, covariant equations without any dependence on a metric or affine structure of spacetime. 

This pre-metric approach can be exclusively derivable from the empirically based postulates of charge and magnetic flux conservation and the Lorentz force (see \cite{Gronwald:2005tv,Hehl:2000pe,Hehl:2005hu,Hehl:1999bt} for a clear axiomatization of electrodynamics). Accordingly, the inhomogeneous equations ($d\bold{G}=\bold{J}$) can be derived from charge conservation ($d\bold{J}=0$) and the homogeneous equations ($d\bold{F}=0$) express magnetic flux conservation. Here, $\bold{F}$ and $\bold{G}$ are the electromagnetic field and the excitation 2-forms, respectively, while $\bold{J}$ is the charge current density 3-form and $d$ stands for exterior derivative. In this formalism, the geometrical and physical interpretations become very simplified and clear. Assuming a 3+1 spacetime splitting (foliation), the Faraday 2-form $\bold{F}$ can be decomposed into an electric part $\bold{E}$, which is a 1-form related to lines, and a magnetic part $\bold{B}$, a 2-form related to surfaces. Similarly, the excitation 2-form $\bold{G}$ contains the electric 2-form and magnetic 1-form excitations, $\bold{D}$ and $\bold{H}$, related to surfaces and lines, respectively. In order for the theory to be complete and to have a predictive power, some form for the constitutive relations between the field strengths [$\bold{F}=(\bold{E},\bold{B})$] and the excitations [$\bold{G}=(\bold{D},\bold{H})$] is required, which constitutes a separate independent postulate in its own. In vacuum, these relations can be viewed as constitutive relations for the spacetime itself and its form will determine the electromagnetic theory that results and its physical predictions (see for example \cite{Hehl:1999bt}). While the field equations rest on empirically well-established postulates, the constitutive relations usually assumed to be local, linear, homogeneous and isotropic have a not so well empirical basis. When considered in vacuum, these relations require the metric structure of spacetime or more specifically, the conformally invariant part of the metric \cite{Hehl:1999bt}. 

One concludes that at the very foundations of electromagnetism the field equations are completely general, without the need of any metric or affine structure of spacetime, but its realization in spacetime via the constitutive relations, reveal a fundamental connection between electrodynamics and the causal (conformal) structure of spacetime. In fact, Friederich Hehl and Yuri Obukhov starting from pre-metric electrodynamics and assuming local and linear constitutive relations, were able to derive the light cone structure and therefore, the conformally invariant part of the metric, provided that there is no birefringence (double refraction) in vacuum \cite{Hehl:2005hu}.
The axiomatic approach developed by Hehl and Obukhov is complementary and compatible with the more traditional Lagrangian formulation \cite{Gronwald:2005tv}. Indeed, the constitutive relations are assumed in the action and the form of these therefore determines the resulting differential field equations. Having in mind the simplifying power of the pre-metric formalism of electrodynamics plus spacetime constitutive relations, in any case, the tensor or components formalism provides a realization of the field equations in spacetime (assuming specific constitutive relations), requiring the metric and affine structure. 

For the case of (pseudo) Riemann geometry, one can then explore the effects of spacetime curvature on electromagnetic phenomena, derived from generalized Gauss and Maxwell-Amp\`{e}re laws and wave equations. Accordingly, the effects of gravity on Maxwell fields, due to the curvature of the background spacetime,  provide a window for the study and testing of theories of gravity (with Riemann geometry) with potential astrophysical applications related to blackholes, pulsars, relativistic stars and gravitaional waves (GW), for example. In this concern, it has been shown that gravitational waves affect the polarization of light (see for example \cite{Hacyan:2015kra}) but it was done in the geometric optics limit and deserves further research. With the advanced LIGO and VIRGO GW detectors (\cite{GWsRiles:2012yw, Singer:2014qca}), a small gravitational signal is expected to be detected within the near future and such an achievement will most probably be celebrated as an important mark of a future new window for astronomy, astrophysics and cosmology (see \cite{GWsRiles:2012yw} for a review on the physics of GW and detectors). Therefore, the intimate relation between gravity, electromagnetism and spacetime geometry should be profoundly explored as it may reveal new alternative approaches for GW detection and also for the study of GW emission by astrophysical sources.

A more consistent study of the coupling between gravity and electromagnetic fields can be achieved through the Einstein-Maxwell coupled equations or similar systems of equations for alternative theories of gravity coupled to electromagnetism. The Einstein-Maxwell system is able to successfully explain many phenomena such as the deflection of light, the gravitational redshift and the Shapiro time delay (see \cite{Efstathiou, Hartle}, for example), but the geometrical optics limit is usually assumed. The exploration of the gravity-electromagnetic coupling beyond that limit continues to be of the utmost importance.   

In this work we will explore the intimate relation between electromagnetism and spacetime geometry and study the observable effects of the curvature (gravity) of Riemann geometry in electromagnetic fields. We will not consider the effect of Maxwell fields on the background geometry in this simplified first approach, which may nevertheless provide some idea of the indeed richer electromagnetic phenomena that derive from such coupling.

This paper is organized in the following manner: In Section \ref{section2}, we review the foundations of electromagnetism based on charge and magnetic flux conservations, the Lorentz force and constitutive relations. The theory follows essentially the approach developed by F. Hehl {\it et al.} \cite{Gronwald:2005tv,Hehl:2000pe,Hehl:2005hu,Hehl:1999bt}. By revisiting the 3-dimensional formulation using integration theory within a 3+1 spacetime splitting (foliation) \cite{Gronwald:2005tv}, this allows to clarify the geometrical interpretations of the electromagnetic quantities and their relations. We then expose the same axiomatic theory in the language of forms in the 3-dimensional and also in the more general 4-dimensional formalism \cite{Hehl:2000pe,Hehl:2005hu,Hehl:1999bt}. We assume local, linear constitutive relations and briefly address the topic on how different forms for these relations can affect the electromagnetic theory and related physical predictions. In Section \ref{section3}, we assume a (pseudo) Riemann manifold for the background spacetime and consider the electromagnetic theory in the tensor formalism assuming the usual local, linear, homogeneous and isotropic constitutive relations. The inhomogeneous equations, wave equations, charge conservation and Lorentz force are explicitly expressed on the curved  spacetime. The generalized Gauss and Maxwell-Amp\`{e}re laws as well as the wave equations reveal a much greater electromagnetic richness with extra terms and electromagnetic couplings induced by the spacetime geometry, with potential interesting applications for different astrophysical scenarios. Finally, in Section \ref{Conclusion}, we summarize and discuss our results.

\section{Foundations of electromagnetism}\label{section2}

Electrodynamics relies on conservation laws and symmetry principles, also known from elementary particle physics. These symmetries are incorporated in the gauge theory and related action principle. Nevertheless, the variational principle is not the unique way to formally derive the electromagnetic theory. In the classical framework, we review the axiomatic approach developed by Friedrich W. Hehl and collaborators  \cite{Gronwald:2005tv,Hehl:2000pe,Hehl:2005hu,Hehl:1999bt} that use specific physical postulates and mathematical methods, namely, the calculus of differential forms but also integration theory, the Poincar\'{e} lemma and the Stokes theorem in the context of tensor analysis in 3-d space. There are two related ways of deriving Maxwell's theory with these tools: the first is based on integration theory and the second on the exterior calculus of differential forms. These approaches make clearer the geometrical significance of the fundamental electromagnetic quantities and their relations. Both methods rely on four basic physical principles or postulates. In the language of forms, the first three axioms enable to express electrodynamics in a pre-metric way. We will also present the 4-dimensional formalism using forms in which the most general field equations are completely pre-metric, coordinate-free and covariant. The 3-dimensional representation is based on a foliation of the spacetime manifold, requiring a certain choice for a (3+1) splitting  in spatial hypersurfaces and an orthogonal time direction.

The starting point for a formal derivation of Maxwell's theory comes in the form of the following four main axioms (postulates):
 
 \begin{itemize}                
    \item Axiom 1: Charge conservation;
    \item Axiom 2: Magnetic flux conservation;
    \item Axiom 3: Lorentz force;
    \item Axiom 4: Linear constitutive (spacetime) relations. 
 \end{itemize}
 
These axioms allow us to obtain the principal aspects of the theory (here the ordering of the axioms of magnetic flux conservation and Lorentz force  is interchanged, with respect to the one used by Hehl {\it et al.}). Charge conservation alone is the foundation for the inhomogeneous equations, the Gauss and Maxwell-Amp\`{e}re laws. The homogeneous equations are derivable from magnetic flux conservation and the Lorentz force. The fourth axiom brings in the metric of spacetime, exposing clearly that there is an intimate connection between electromagnetism and spacetime geometry already at the foundational basis of classical electrodynamics. 

Two additional axioms which we will not consider in the present work, related to the energy-momentum distribution of the electromagnetic field \citep{Hehl:2008sm}, are required for a macroscopic description of electromagnetism (in matter). These are the following:

 \begin{itemize}        
    \item Axiom 5: Specification of the energy-momentum distribution of the electromagnetic field by means of the energy-momentum tensor (From this the energy density and the energy flux density, i.e., the Poynting vector, are obtained);
    
    \item Axiom 6: Splitting of the total electric charge and currents in a bound or material component which is conserved and a free or external component. 
 \end{itemize}

Throughout this work, indices with Greek letters represent spacetime indices ranging from 0 to 3, while Latin indices are spatial from 1 to 3. Unless stated otherwise, repeated indices imply Einstein's summation convention and we will be adopting a $(+---)$ signature for the spacetime metric. 

\subsection{Axiomatic structure of Maxwell theory: Pre-metric approach plus constitutive (spacetime) relations}

\subsubsection{3-d formalism: Geometrical and physical interpretations using integration theory and differential forms}

An important link between integration theory and geometrical considerations follows from the fact that integration
is required to yield invariant quantities under arbitrary coordinate transformations. In 3-dimensional space there are three basic  geometrical possibilities for integration, i.e., along lines, 2-surfaces and volumes. Taking into consideration the way in which line, surface and volume elements transform under general coordinate transformations, it is possible to find the correct or natural line, surface and volume integrands which transform in a complementary way in order to give (geometricaly) invariant results. One concludes that:
 \begin{itemize}        
    \item Covectors (1-forms) $\alpha_{i}$ are natural line integrands;
    
    \item Vector densities $\beta^{j}$ are natural surface integrands;
    
    \item Scalar densities $\varrho$ are natural volume integrands.
 \end{itemize}
 
These quantities transform under arbitrary changes of coordinates $x^{i}\rightarrow x^{j'}$ according to
\begin{equation}
\alpha_{k'}=\alpha_{i}\partial_{k'}x^{i},\quad \beta^{j'}=\vert J\vert^{-1}\beta^{i}\partial_{i}x^{j'}, \quad \varrho'=\vert J\vert^{-1}\varrho,
\end{equation}
where $J={\rm det}\left[\partial x^{k'}/\partial x^{j}\right]$ is the Jacobian of the coordinate transformation and $\vert J\vert=\sqrt{\vert g^{3d}\vert}$, where $g^{3d}$ is the determinant of the 3-dimensional metric. Note that if one considers the Jacobian $J$  in the above expressions instead of its modulus, the scalar and vector densities can change sign under reflections. In this case they are sometimes designated by pseudo-tensor densities. 

The Poincar\'{e} lemma states under which conditions certain mathematical objects can be expressed in terms of derivatives of other objects (potentials). Consider the natural integrands $\alpha_{i}$, $\beta^{j}$, $\varrho$ of line, surface and volume integrals respectively. Let us assume that they are defined in open connected regions of 3-dimensional space. Then:
\begin{enumerate}

\item If $\alpha_{i}$ is curl free, it can be written as the gradient of a scalar function $f$,
\begin{equation}
\epsilon^{ijk}\partial_{j}\alpha_{k}=0 \quad \Rightarrow \quad \alpha_{k}=\partial_{k}f .
\end{equation}

 \item If $\beta^{j}$ is divergence free, it can be written as the curl of the integrand $\alpha_{i}$ of a line integral,
\begin{equation}
\partial_{i}\beta^{i}=0 \quad  \Rightarrow \quad  \beta^{i}=\epsilon^{ijk}\partial_{j}\alpha_{k} .
\end{equation}
 
 \item The integrand $\varrho$ of a volume integral (scalar density) can be written as the divergence of an integrand $\beta^{i}$ of a surface integral (vector density),
\begin{equation}
\varrho=\partial_{i}\beta^{i}.
\end{equation}
 
    \end{enumerate}

In the expressions above, $\epsilon^{ijk}$ corresponds to the completely antisymmetric Levi-Civita (pseudo) tensor.
We are now ready to proceed with the physical postulates underlying the electromagnetic theory.\\
 
\paragraph{Charge conservation and Maxwell's inhomogeneous equations:}

With these mathematical methods it follows that charge densities as natural volume integrands are scalar densities $\varrho$ and current densities as natural surface integrands are vector densities $\jmath^{k}$ and the postulate of charge conservation
\begin{equation}
\partial_{t}\varrho+\partial_{i}\jmath^{i}=0,
\label{chargecons}
\end{equation}
allows us to define the electric and magnetic excitations as natural surface and line elements respectively, obeying the inhomogeneous Maxwell equations. In fact, using the Poincar\'{e} lemma it follows that
\begin{equation}
\varrho=\partial_{i}D^{i},\label{Gauss} 
\end{equation}
where $D^{i}$ is a vector density, and therefore related to surfaces, thus deriving the Gauss law. On the other hand, substituting the above expression on charge conservation and using the Poincar\'{e} lemma we deduce
\begin{equation}
\partial_{i}\left(\partial_{t}D^{i}+\jmath^{i} \right)=0 \quad \Rightarrow  \quad \partial_{t}D^{i}+\jmath^{i}=\epsilon^{ijk}\partial_{j}H_{k},\label{Maxamp}
\end{equation}
where $H_{k}$ is a line integrand and the Maxwell-Amp\`{e}re law is thus derived.

It should be clear that we have obtained the inhomogeneous equations and the electric and magnetic excitations from charge conservation and the Poincar\'{e} lemma without introducing the concept of force. Notice that since charge conservation is valid in microscopic physics, the same is true for the inhomogeneous equations and for the excitations, contrary to what is commonly stated in the literature.

It should be said that the notations used above for the gradient, divergence and curl were used simbolicaly. For the (3+1) splitting of general (pseudo) Riemanian spacetime, all spatial derivatives must be replaced by covariant derivatives and a fully 4-dimensional (covariant) approach can be obtained from Eqs. (\ref{chargecons})-- (\ref{Maxamp}) replacing all derivatives (spatial and temporal) by covariant ones.\\

\paragraph{Magnetic flux conservation, Lorentz force and Maxwell's homogeneous equations:}

There is some analogy between vortex lines in hydrodynamics and magnetic flux lines. Helmholtz' works on hydrodynamics enabled to conclude that vortex lines are conserved. Vortex lines that pierce a 2-surface can be integrated over to originate a scalar called circulation. Circulation in a perfect fluid is constant provided the loop enclosing the surface moves along with the fluid. Analogously, there is good experimental evidence that magnetic flux is conserved. In fact, it seems that at the microscopic level magnetic flux occurs in quanta and the corresponding magnetic flux unit is called flux quantum or fluxon. One fluxon carries $\Phi_{0}=h/2e\approx 2,7\times10^{-15} Wb$ , where $e$ is the elementary charge and $h$ is Planck's constant \cite{HehlYuribook}. Single quantized magnetic flux lines have been observed in the interior of type II superconductors when exposed to sufficiently strong magnetic fields (\cite{HehlYuribook}, p. 131) and they can be counted.
 
We therefore assume that magnetic flux, defined as
\begin{equation}
\Phi_{mag}\equiv\int\int B^{i}da_{i},
\end{equation}
is conserved, where the magnetic field $B^{i}$ is a natural surface integrand and therefore, a vector density. The corresponding continuity equation (analogous to charge conservation)
\begin{equation}
\partial_{t}\Phi_{mag}+\oint j^{\Phi}_{i}dx^{i}=0,
\end{equation}
allows us to define the magnetic flux current density $j^{\Phi}_{i}$ as a natural line integrand (covector). Applying Stokes' theorem
\begin{equation}
\oint j^{\Phi}_{i}dx^{i}=\int\int \epsilon^{ijk}\partial_{j}j^{\Phi}_{k}da_{i} ,
\end{equation}
locally we get,
\begin{equation}
\partial_{t}B^{i}+\epsilon^{ijk}\partial_{j}j^{\Phi}_{k}=0.
\end{equation}

On the other hand, force is integrated through lines to yield work, therefore force is a natural line integrand, i.e., a covector $f_{i}$. The Lorentz force postulate  
\begin{equation}
f_{i}=q(E_{i}+\epsilon_{ijk}u^{j}B^{k}),
\end{equation}
implies that the electric field is also a line integrand, i.e., a covector $E_{k}$. Now, since $j^{\Phi}_{k}$ and the electric field have the same physical dimensions (in S.I. units, flux/(time$\times$length) correspond to $V/m$) and geometrical properties, it is plausible to make the identification $j^{\Phi}_{k}=E_{k}$ (in accordance to the Lenz rule) and recover the Faraday law, which expresses magnetic flux conservation
\begin{equation}
\partial_{t}B^{i}+\epsilon^{ijk}\partial_{j}E_{k}=0.
\end{equation}
Taking the divergence of this expression (remembering that the divergence of a curl is zero), we obtain 
\begin{equation}
\partial_{i}(\partial_{t}B^{i})=0,
\end{equation}
and taking into account the Poincar\'{e} lemma, we can define the magnetic charge scalar density $\varrho_{mag}\equiv \partial_{i}B^{i}$, and therefore conclude that
\begin{equation}
\partial_{t}\varrho_{mag}=0,
\end{equation}
which rigorously states that the magnetic charge must be static in accordance to magnetic flux conservation.
Now, since $\varrho_{mag}$ is a scalar density, under a general coordinate transformation $\lbrace x,t\rbrace\rightarrow \lbrace x',t'\rbrace$, it transforms according to $\varrho_{mag}'=\vert J^{-1}\vert\varrho_{mag}$. Therefore, in general, $\partial_{t'}\varrho_{mag}'\neq0$ and so we set it to zero, i.e.,
\begin{equation}
\varrho_{mag}=0 \quad  \Rightarrow \quad   \partial_{i}B^{i}=0,
\end{equation}
which expresses the Gauss law for magnetism and the absence of magnetic monopoles. In other words, magnetic flux conservation is incompatible with magnetic monopoles.

We conclude that the electric field $E$ and magnetic excitation $H$ are both related to lines while the magnetic field $B$ and electric excitation $D$ are both related to surfaces, i.e.,
\begin{itemize}

\item
Electric field: 1-form (covector); line integrand, $E_{i}$;

\item
Magnetic field: vector density; surface integrand, $B^{j}$;

\item
Electric excitation: vector density; surface integrand, $D^{j}$;

\item
Magnetic excitation: 1-form (covector); line integrand, $H_{i}$.

\end{itemize}

\paragraph{Constitutive relations:}
Maxwell's equations constitute 8 equations (6 of which are dynamical) with 12 unknown quantities. In order to solve these equations we need to postulate a form for the so-called constitutive relations, $D=D(E,B)$ and $H=H(E,B)$, between the excitations and the electric and magnetic fields. With these relations, Maxwell's equations reveal two independent electromagnetic degrees of freedom. The solutions to these equations and all the associated electromagnetic phenomena depend crucially on our assumption regarding the nature of the constitutive relations.

In order to relate the electromagnetic fields and their excitations there are two points to consider. One is geometrical and the other is physical. First, the physical consideration has to do with the dimensions involved and it is at this level that the electric  permittivity and magnetic permeability tensors are introduced, characterizing the material medium or empty space. These so-called electromagnetic properties of vacuum are usually assumed to be homogeneous and isotropic, represented by diagonal matrices with equal constant diagonal components. This assumption is not necessarily the unique choice and one could argue that the permittivity and permeability tensors reflect electromagnetic properties of spacetime (or of electro-vacuum) and should reflect the spacetime symmetries. We will come back to this point.  In second place, geometrically, the constitutive relations imply that one needs to relate 1-forms (co-vectors) to vector densities (which can be mapped to 2-forms). It follows that the spacetime metric enables to realize the required link. In fact, $g^{ij}\sqrt{\vert g\vert}$ transforms like a density and maps a covector (1-form) to a vector density.

With these considerations, we will assume local, linear, homogeneous and isotropic constitutive relations in vacuum without mixing electric and magnetic properties, through the following expressions
\begin{equation}
D^{j}=\varepsilon_{0}\sqrt{\vert h\vert}h^{ij}E_{i}, \quad 
E_{i}=\dfrac{1}{\varepsilon_{0}\sqrt{\vert h\vert}}D^{j}h_{ij},
\label{constitutive3delectric}
\end{equation} 
\begin{equation}
B^{j}=\dfrac{\mu_{0}}{\sqrt{\vert h\vert}}h^{ij}H_{i}, \quad 
H_{i}=\dfrac{\sqrt{\vert h\vert}}{\mu_{0}}B^{j}h_{ij}, 
\label{constitutive3dmagnetic}
\end{equation} 
where $h_{ij}$ is the 3-dimensional metric induced on the 3-dimensional hypersurfaces.

The postulates of charge conservation, magnetic flux conservation, Lorentz force and constitutive relations can be clearly expressed using forms leading to the same fundamental geometrical and physical conclusions. In this formalism, charge density $\bold{\rho}$ is a 3-form, current density $\bold{j}$ a 2-form, electric fields $\bold{E}$ are 1-forms (related to lines), magnetic fields $\bold{B}$ are 2-forms (related to surfaces), the electric excitation $\bold{D}$ is a 2-form and the magnetic excitation $\bold{H}$ is a 1-form. 

Considering a (3+1) foliation of  spacetime, the electromagnetic quantities are defined on the 3-dimensional hypersurfaces. For a given foliation, the set of Maxwell's equations
\begin{eqnarray}
d\bold{D}=\bold{\varrho},\qquad d\bold{H}=\bold{j}+\partial_{t}\bold{D},
    \\
 d\bold{E}+\partial_{t}\bold{B}=0, \qquad d\bold{B}=0,\label{maxwelleq3dforms}
\end{eqnarray}
are fully general pre-metric, covariant equations, coming directly from charge conservation, magnetic flux conservation and the Lorentz force. As previously mentioned, to solve this set of equations one requires the (spacetime) constitutive relations relating the electric and magnetic fields to the excitations.
Assuming linear, homogeneous and isotropic constitutive relations, without mixing electric and magnetic properties, these relations, in the language of forms are achieved via the Hodge star operator in 3-dimensional space which maps $k$-forms to ($d-k$)-forms (where $d$ is the dimension of the manifold under consideration), and are given by
\begin{equation}
\bold{D}=\varepsilon_{0}\star\bold{E},\qquad
\bold{H}=\mu_{0}^{-1}\star\bold{B},\label{constitutive3d}
\end{equation}
which introduces the spacetime metric
\begin{eqnarray}
D_{jk}=\varepsilon_{0}\sqrt{\vert h\vert}\epsilon_{ijk}h^{im}E_{m},
\end{eqnarray}
\begin{eqnarray}
 E_{i}=\dfrac{1}{2\varepsilon_{0}\sqrt{\vert h\vert}}\epsilon_{ijk}D_{mn}g^{mj}g^{nk},
\end{eqnarray} 
\begin{eqnarray}
B_{jk}=\dfrac{\mu_{0}}{\sqrt{\vert h\vert}}\epsilon_{ijk}h^{im}H_{m},
\end{eqnarray}
\begin{eqnarray}
H_{i}=\dfrac{\sqrt{\vert h\vert}}{2\mu_{0}}\epsilon_{ijk}B_{mn}g^{mj}g^{nk}.
\end{eqnarray}
  
With this choice, the electric and magnetic excitations are odd forms which expresses their behaviour under spatial reflections \citep{Hehl:2000pe}.
We will come back to these important relations in the 4-dimensional formalism and in the final section of this work since, as previously mentioned, they reveal a fundamental connection between electromagnetic fields, the electromagnetic properties of vacuum and the metric and conformal (causal) structure of spacetime.\par 
 
The 3-dimensional formalism presented here using integration theory and linear forms is completely self-compatible, revealing in a clear way the geometrical meanings implicit to the electromagnetic quantities and their relations. In particular, we can map the electromagnetic 2-forms to the associated vector densities according to
\begin{equation}
D^{a}=\dfrac{1}{2}\epsilon^{abc}D_{bc},\quad D_{ab}=\epsilon_{abc}D^{c},
\end{equation} 
\begin{equation}
B^{a}=\dfrac{1}{2}\epsilon^{abc}B_{bc},\quad B_{ab}=\epsilon_{abc}B^{c}.
\end{equation} 
With the introduction of the constitutive relations, the axiomatic approach to classical electrodynamics is completed.

\subsubsection{4-dimensional formalism using differential forms}

\paragraph{Charge conservation and the inhomogeneous equations:}

In the 4-dimensional formalism, charge conservation can be expressed by saying that the total (net) flux of electric charge through any closed 3-surface is zero. In order to integrate along a 3-surface we then require a 3-form electric charge current density $\bold{J}$, which is related to the usual 4-current vector $j^{\lambda}$ via the Hodge star product of the corresponding 1-form $\bold{j}=j_{\alpha} \bold{dx^{\alpha}}$
\begin{equation}
\bold{J}=\star\bold{j},\qquad 
\bold{J}=\dfrac{1}{3!}\epsilon_{\alpha\beta\gamma\lambda}j^{\lambda}\sqrt{-g}\bold{dx^{\alpha}}\wedge\bold{dx^{\beta}}\wedge\bold{dx^{\gamma}},
\end{equation}
therefore
\begin{eqnarray}
j_{123}=j^{0}\sqrt{-g}=\rho c\sqrt{-g},\qquad j_{230}=j^{1}\sqrt{-g}, 
  \nonumber \\
 j_{301}=j^{2}\sqrt{-g},\qquad j_{012}=j^{3}\sqrt{-g},
 \nonumber
\end{eqnarray}
and $\jmath^{\lambda}\equiv\sqrt{-g}j^{\lambda}$ is a vector density.
We can then write
\begin{equation}
\oint_{3d} {\bold{J}}=\int_{4d} {d\bold{J}}=0,
\end{equation}
where we have applied the fundamental theorem of the exterior calculus of forms, namely, the Stokes theorem. 
The second equality is valid for any compact 4-dimensional volume enclosed by the 3-surface. Therefore we arrive at $d\bold{J}=0$, which expresses charge conservation locally.
Now, since $\bold{J}$ is a 3-form and $dd=0$,  it can be expressed by the exterior differential of a 2-form
\begin{equation}
d\bold{J}=0 \quad  \Rightarrow \quad  d\bold{G}=\bold{J}.\quad 
\end{equation}
Therefore, in the language of forms, it is clear that charge conservation is at the foundation of Maxwell's inhomogeneous equations 
which in this formalism are fully general, coordinate-free, pre-metric and covariant equations.

In component form we have
\begin{equation}
\nabla_{[\alpha}G_{\beta\gamma]}=\epsilon_{\alpha\beta\gamma\lambda}\jmath^{\lambda},
\end{equation}
where $\nabla_\alpha$ represents the covariant derivation defined on the spacetime manifold endowed with affine and metric structure. Therefore, with the following definitions
\begin{equation}
G_{0i}\equiv-H_{i}, \qquad G_{ij}\equiv\epsilon_{ijk}D^{k}c=D_{ij}c,
\end{equation}
the most general expressions for the Gauss and Maxwell-Amp\`{e}re laws in component form are
\begin{equation}
\nabla_{i}D^{i}=\varrho,\qquad \nabla_{0}D^{k}c+\epsilon^{ijk}\nabla_{i}H_{j}=\jmath^{k},\label{generalinhom}
\end{equation}
where $\varrho\equiv\sqrt{-g}\rho$ and $\rho$ is the charge density.
Indeed, assuming the validity of a local foliation of spacetime, the electromagnetic excitation 2-form can be written in terms of its spatial and temporal parts establishing a link with the 3-dimensional formalism previously discussed
\begin{equation}
\bold{G}=\bold{H}\wedge dx^{0}+\bold{D}c.
\end{equation}

\vspace{2mm}

\paragraph{Magnetic flux conservation and the homogeneous equations:}
Magnetic flux conservation can be expressed by
\begin{equation}
\oint_{surface}\bold{F}=0 \quad  \Rightarrow \quad  \int_{volume} d\bold{F}=0,
\end{equation}
where $\bold{F}$ is the Faraday 2-form, obeying the homogeneous equations
\begin{equation}
d\bold{F}=0 \quad  \Rightarrow  \quad  \bold{F}=d\bold{A},
\label{dfzero}
\end{equation}
and $\bold{A}$ is the electromagnetic potential 1-form. The magnetic flux conservation is at the foundation of the homogeneous equations which also naturally follow as a Bianchi identity, resulting from the derivation of the potential twice. Using the spacetime foliation we can write
\begin{equation}
\bold{F}=dx^{0}\wedge\bold{E}\dfrac{1}{c}-\bold{B}.
\end{equation}

\paragraph{Homogeneous and isotropic constitutive relations:}
In this formalism the linear, local, homogeneous and isotropic constitutive relations can be expressed through the Hodge star operator
\begin{equation}
\bold{G}=\mu_{0}^{-1}\star\bold{F}.\label{constitutive}
\end{equation} 
With this assumption or postulate the inhomogeneous equations can then be written by
\begin{equation}
d(\star\bold{F})=\mu_{0}\bold{J},
\end{equation}
or in terms of the potential by
\begin{equation}
d\star d\bold{A}=\mu_{0}\bold{J}.
\end{equation}
In component form the above constitutive relations are
\begin{equation}
G_{\mu\nu}=\dfrac{1}{2\mu_{0}}\sqrt{-g}g^{\alpha\lambda}g^{\beta\gamma}\epsilon_{\mu\nu\lambda\gamma}F_{\alpha\beta}.\label{constitutive4d}
\end{equation}
The factor $\sqrt{-g}g^{\alpha\lambda}g^{\beta\gamma}\epsilon_{\mu\nu\lambda\gamma}$ is conformally invariant. Therefore, one arrives at
\begin{eqnarray}
D^{j}&=&\sqrt{-g}\Big[ \varepsilon_{0}E_{k}\left(g^{0j}g^{k0}-g^{kj}g^{00}\right)
    \nonumber \\
&&-c^{-1}\mu_{0}^{-1}\dfrac{1}{2}B_{mn}(g^{mj}g^{n0}-g^{m0}g^{nj})\Big],\label{contitutD}  
\end{eqnarray}
\begin{eqnarray}
\hspace{-0.82cm}  H_{k}=\sqrt{-g}\mu_{0}^{-1}\Bigg[\dfrac{1}{2}B_{ij}\epsilon_{krs}g^{ir}g^{js}
-c^{-1}E_{j}\epsilon_{krs}g^{0r}g^{js}\Big].
\end{eqnarray}
It is clear that assuming linear constitutive relations of the form (\ref{constitutive}), we are not excluding a mixing between electric and magnetic quantities, in contrast to the expressions in Eq. (\ref{constitutive3d}). In fact, according to the expressions above, this mixing will occur whenever the metric has off-diagonal elements involving the time-space components.
Notice that, in general $-g=g_{00}h$ and for the (3+1) splitting of space+time we recover the constitutive relations in Eqs. (\ref{constitutive3delectric}) and (\ref{constitutive3dmagnetic}). For a diagonal metric, we have
\begin{equation}
D^{j}=-\sqrt{-g}\varepsilon_{0}E_{j}g^{jj}g^{00},  
\end{equation}
\begin{equation}
H_{k}=\sqrt{-g}\mu_{0}^{-1}B^{k}g_{kk},  
\end{equation}
where no contraction (summation rule) is assumed on the above expressions and we have used the fact that $\epsilon_{krs}\epsilon^{rsf}=2\delta^{f}_{k}$.

In Minkowski spacetime we get the familiar relations in vacuum, which assume homogeneity and isotropy.\\

\paragraph{Action principle:}
Maxwell's equations for the fields $E$ and $B$ can be derived from the following 4-form
\begin{equation}
\bold{S}=\int{\bold{F}\wedge\bold{G}}+\int{\bold{J}\wedge\bold{A}},\label{action}
\end{equation}
assuming a specific set of constitutive relations between $\bold{G}=(\bold{H},\bold{D})$ and $\bold{F}=(\bold{E},\bold{B})$. For the homogeneous and isotropic linear constitutive relations in Eq. (\ref{constitutive}) we get the usual free field action of electromagnetism,
\begin{equation}
\bold{S}_{free}=\dfrac{1}{\mu_{0}}\int{\bold{F}\wedge\star\bold{F}},\label{usualaction}
\end{equation}
normally presented in relation to the gauge approach. It is clear that the constitutive relations (which imply the metric structure of spacetime) and the form of these relations are implicit in the usual (gauge approach)  inhomogeneous equations.

\subsection{More general linear constitutive relations}

The Maxwell equations together with the spacetime relations, constitute the foundations of classical electrodynamics. These laws, in the classical domain, are assumed to be of universal validity. Only if vacuum polarization effects of quantum electrodynamics are taken into account or hypothetical non-local terms should emerge from huge accelerations, axiom 4 can pick up corrections yielding a nonlinear law (Heisenberg-Euler electrodynamics \cite{HeisenbergEuler}) or a nonlocal law (Volterra-Mashhoon electrodynamics \cite{Mashhoon}), respectively. In this sense, the Maxwell equations are more general than the constitutive spacetime relations, however, the latter are not completely untouchable. We may consider them as constitutive relations for spacetime itself, as discussed below.

As previously mentioned, the constitutive relations in vacuum not only introduce the spacetime metric but also the vacuum electromagnetic properties via the electric permitivity and magnetic permeability tensors. The assumption of homogeneous and isotropic relations is based on the assumption that these vacuum electromagnetic properties are homogeneous and isotropic. Can we drop these assumptions? It is clear that if only homogeneity is abandoned, then the velocity of light in vacuum can in principle vary with the spacetime point without loosing local Lorentz invariance. Even more generally, the principle of (local) conformal invariance, which guarantees the invariance of the casual structure of spacetime (locally), does not require homogeneity or isotropy for the speed of light in vacuum. As previously said, one argument in favour of letting go of the assumption of homogeneity and isotropy for the electromagnetic properties of vacuum is that these quantities could characterize the ``electro-vacuum'' which can be intimately related to spacetime geometry and therefore to its symmetry properties. In this sense, it seems more natural to assume that the symmetry properties of the tensors $\varepsilon_{ij}$ and $\mu_{km}$ in vacuum follow the spacetime isometries. 

This reasoning could come from a self-compatible interpretation of the coupled Einstein-Maxwell equations. Electromagnetic fields affect spacetime geometry and this geometry affects the propagation of the fields. In fact, in the spirit of general relativity, the metric is not {\it a priori} given, it depends on the local energy-momentum content of physical fields. Therefore, spacetime symmetries are also not {\it a priori} given, they must be considered locally for each physical scenario. Why should the properties of vacuum, such as the electric permitivity and magnetic permeability be {\it a priori} given, in particular, why should these be homogeneous and isotropic for axially or spherically symmetric spacetime? For example, in spherical symmetric cases like the Schwarzschild solution, according to the interpretation here proposed, the speed of light in vacuum could have a dependence with the radial coordinate and this result could be tested experimentally.

A very simple expression for the linear constitutive relations (in vacuum) assuming a local (3+1) foliation can be given by
\begin{equation}
D^{i}=\sqrt{\vert h\vert}(\varepsilon_{0})^{ij}E_{j},\qquad H_{i}= \sqrt{\vert h\vert}
(\mu_{0}^{-1})_{ij}
B^{j}.\label{constpermitiv}
\end{equation} 
With these expressions we are assuming locality, linearity and a non-mixing between electric and magnetic components but without forcing the assumptions of homogeneity and isotropy. These relations will affect the inhomogeneous equations (\ref{generalinhom}). In particular, for physical conditions where the spacetime metric has spherical symmetry, according to the interpretation here suggested, the permittivity and permeability tensors follow the spacetime isometries and therefore become diagonal with equal components (isotropy) but with a radial dependence on position (inhomogeneity and spherical symmetry), i.e., $(\varepsilon_{0})^{j}_{k}=\varepsilon_{0}(r)\delta^{j}_{k}$, $(\mu_{0})^{j}_{k}=\mu_{0}(r)\delta^{j}_{k}$. When $(\varepsilon_{0})^{j}_{k}=\varepsilon_{0}\delta^{j}_{k}$ and $(\mu_{0})^{j}_{k}=\mu_{0}\delta^{j}_{k}$ we recover the homogeneous and isotropic relations.

Following the approach of Hehl and Obukhov \cite{Hehl:1999bt,HehlYuribook}, the most general expression for linear (local) relations in the 4-dimensional formalism is the following
\begin{equation}
G_{\mu\nu}=\chi_{\mu\nu}^{\alpha\beta}F_{\alpha\beta}=\chi_{\mu\nu}^{[\alpha\beta]}F_{\alpha\beta},\label{consttensor}
\end{equation}
where the tensor $\chi_{\mu\nu}^{\alpha\beta}$ is antisymmetric in the lower indices and has, in general, 36 independent components. This tensor can be decomposed into its irreducible components where the principal part is related to the relations in Eq. (\ref{constitutive4d}).
The constitutive equations in matter are more complicated (see \citep{HehlYuribook,Baekler:2014kha,Hehl:2004tk}) and it would be appropriate to derive them, using an averaging procedure, from a microscopic model of matter. For instance, this lies within the subject of solid state or plasma physics. Hehl and Obukhov arrived at the following relations for a general linear magnetoelectric medium \citep{Hehl:2005hu}
\begin{equation}
D^{i}=\left(\varepsilon^{ij}-\epsilon^{ijk}n_{k}\right)E_{j}+\left(\gamma_{\,j}^{i}+\tilde{s}_{j}^{\,i}
\right)B^{j}+
\left(\alpha-s
\right)\delta^{i}_{j}B^{j},
\end{equation} 
\begin{equation}
H_{i}=\left(
\mu^{-1}_{ij}
-\epsilon_{ijk}m^{k}\right)B^{j}+
\left(-\gamma_{\,i}^{j}+\tilde{s}_{i}^{\,j}
\right)E_{j}+
\left(\alpha+s
\right)\delta^{j}_{i}E_{j},
\end{equation} 
where the matrices $\varepsilon_{ij}$ and $\mu_{ij}^{-1}$  are symmetric and have 6 independent components each. They correspond to the permittivity and impermeability tensors (reciprocal permeability tensor).
The magnetoelectric cross-term $\gamma_{ij}$, which is trace-free ($\gamma^{j}_{j}=0$), has 8 independent components. It is related to the Fresnel-Fizeau effects \cite{Fizeau}. The 4-dimensional pseudo-scalar $\alpha$, called axion, corresponds to the perfect electromagnetic conductor of Lindell and Sihvola \citep{LindellShivola}, a Tellegen type structure \citep{Tellegen}. Until now, we considered a total of 6+6+8+1=21 independent components. We can have 15 more components related to dissipation (which cannot be derived from a Lagrangian). The 3+3 components of $n_{k}$ and $m_{k}$ (electric and magnetic Faraday effects), 8 components from the matrix   $\tilde{s}_{ij}$ (optical activity), which is traceless and 1 component from the 3-dimensional scalar $s$ (spatially isotropic optical activity).

We end up with a general linear medium with 20+1+15=36 independent components. Notice that we didn't include the $\sqrt{-g}$ factor in the permitivity and permeability terms which is one in Minkowski spacetime, but it should be present in the general Riemannian case. 

These constitutive relations can also be applied to vacuum with linear electromagnetic properties. This topic requires further investigation since these relations might be viewed as relations for spacetime itself which would imply a deep connection between physical properties of (classical) vacuum and spacetime (suggesting or reinforcing the idea of spacetime physicalism, i.e., spacetime with well defined physical ontology).

From a variational point of view the equations for the permittivity and permeability tensors in Eq. (\ref{constpermitiv}) or for the tensor $\chi_{\mu\nu}^{\alpha\beta}$ in Eq. (\ref{consttensor}), could in principle be obtained from an appropriate action corresponding to a tensor-vector electromagnetic theory, as long as dissipation effects are disregarded.

\section{Electrodynamics in curved spacetime}\label{section3}

\subsection{Field equations}

\subsubsection{The background spacetime: (pseudo)Riemannian geometry}

Consider a 4-dimensional spacetime manifold with pseudo-Riemannian geometry. The metric $g_{\alpha\beta}$ is the fundamental object required to compute spacetime distances and the connection $\Gamma^{\alpha}_{\;\beta\gamma}$ is the fundamental object required to define covariant differentiation $\nabla_{\mu}$ and therefore parallel transport. In the (pseudo) Riemann geometry assumed in general relativity the connection is metric compatible, i.e.,
\begin{equation}
\nabla_{\mu}g^{\alpha\beta}=0,\label{metriccompatible}
\end{equation}
which implies the invariance of the inner product of vectors under parallel transport.
The only symmetric connection obeying this condition is the so-called Levi-Civitta connection which is not independent from the metric according to the well known relation
\begin{equation}
\Gamma^{\alpha}_{\;\mu\nu}=g^{\alpha\lambda}(\partial_{\nu}g_{\lambda\mu}+\partial_{\mu}g_{\lambda\nu}-\partial_{\lambda}g_{\mu\nu}).\label{levicivitta}
\end{equation}
The symmetry of the connection implies that the manifold is torsionless. As a consequence, the auto-parallel geodesics coincide with extremal geodesics. The first are obtained by requiring parallel transport of the tangent vector to the curve, while the second comes from extremizing the spacetime distance between two points along the curve. The spacetime line element is $ds^{2}=g_{\alpha\beta}dx^{\alpha}dx^{\beta}$.
%
%
We recall that we are adopting a $(+---)$ signature, and Greek letters are spacetime indices ranging from 0 to 3 while Latin indices are space indices from 1 to 3.

\subsubsection{Maxwell's inhomogeneous equations}

We will consider electromagnetic fields on this curved spacetime background without taking into consideration the influence of the electromagnetic fields on spacetime geometry. To explore the physical applications we will consider the usual field equations derived from the action principle, which assume implicitly local, linear, homogeneous and isotropic constitutive relations (\ref{constitutive4d}).  

The Faraday tensor previously introduced in Eq. (\ref{dfzero}) has the following components
\begin{equation}
F_{\mu\nu}=\nabla_{\mu}A_{\nu}-\nabla_{\nu}A_{\mu}=\partial_{\mu}A_{\nu}-\partial_{\nu}A_{\mu},\label{Faraday}
\end{equation}
where the last equality follows from the fact that the spacetime manifold is torsionless, i.e., the anti-symmetric part of the connection is zero. Note that following the formalism introduced in the first section, $F_{\mu\nu}$ are the components of the Faraday 2-form $\bold{F}$ and $A_{\lambda}$ are the components of the electromagnetic potential 1-form $\bold{A}$. On the other hand, we saw that the electric field $\bold{E}$ is related to lines being a natural line integrand and therefore is represented by a 1-form (a covector) in the 3-dimensional space, while the magnetic field $\bold{B}$ is related to surfaces being represented by a 2-form with components $B_{jk}$ that can be mapped to the (contravariant) components of a vector density $B^{i}$ which are natural surface integrands. Therefore, we introduce the following definitions 
\begin{eqnarray}
F_{0k}&=&\dfrac{1}{c}\partial_{t}A_{k}-\partial_{k}A_{0}\equiv \dfrac{E_{k}}{c},
   \\
 F_{jk}&=&\partial_{j}A_{k}-\partial_{k}A_{j}\equiv -B_{jk}=-\epsilon_{ijk}B^{i}.\label{magneticfield}
\end{eqnarray}
The spacetime metric and its inverse are used to lower or raise indices and so we have
\begin{equation}
F^{\mu\nu}=g^{\alpha\mu}g^{\beta\nu}F_{\alpha\beta},\quad A^{\mu}=g^{\mu\nu}A_{\nu}.\label{upperindF}
\end{equation}

The well-known Maxwell equations, which follow from the action in Eq. (\ref{usualaction}) (with the appropriate source term) and from a Bianchi identity (\ref{dfzero}), and given by
\begin{equation}
\nabla_{\mu}F^{\mu\nu}=\mu_{0}j^{\nu},\label{Maxeq}\quad \nabla_{[\alpha}F_{\beta\gamma]}=0,
\end{equation}
are compatible with charge conservation and magnetic flux conservation. Very importantly, the inhomogeneous equations above implicitly assume local, linear, homogeneous and isotropic constitutive relations. If we change these relations, we get a different set of field equations, with new predictions. This set of 8 equations includes the influence of spacetime geometry coming from the covariant derivative
\begin{equation}
\nabla_{\mu}F^{\lambda\nu}=\partial_{\mu}F^{\lambda\nu}+\Gamma^{\lambda}_{\alpha\mu}F^{\alpha\nu}+\Gamma^{\nu}_{\alpha\mu}F^{\lambda\alpha},\label{covar}
\end{equation}
and from the fact that the (inverse) metric is used to raise indices according to Eq. (\ref{upperindF}). Recall that the general expression for the divergence of an antisymmetric tensor $\Theta^{\alpha\beta}$ in (pseudo) Riemann spacetime is given by
\begin{equation}
\nabla_{\mu}\Theta^{\mu\nu}=\dfrac{1}{\sqrt{-g}}\partial_{\mu}\left( \sqrt{-g}\Theta^{\mu\nu}\right). 
\end{equation}
In fact, after contracting $\mu$ with $\lambda$, the third term in Eq. (\ref{covar}) vanishes, since the (symmetric) connection is contracted with the components of an anti-symmetric tensor and on the other hand, for (pseudo) Riemann manifolds we can use the following relation $\Gamma^{\varepsilon}_{k\varepsilon}=\partial_{k}\left( \log (\sqrt{-g})\right)$, where a contraction is assumed in $\varepsilon$. Therefore, the inhomogeneous equations are given by
\begin{equation}
\partial_{\mu}F^{\mu\nu}+\dfrac{1}{\sqrt{-g}}\partial_{\mu}(\sqrt{-g})F^{\mu\nu}=\mu_{0}j^{\nu},\label{Maxeqriem1}
\end{equation}
therefore,
\begin{eqnarray}
g^{\alpha\mu}g^{\beta\nu}\partial_{\mu}F_{\alpha\beta}
+F_{\alpha\beta}\Big[\partial_{\mu}(g^{\beta\nu}g^{\alpha\mu})
   \nonumber  \\
+g^{\alpha\mu}g^{\beta\nu}\dfrac{1}{\sqrt{-g}}\partial_{\mu}(\sqrt{-g})\Big] =\mu_{0}j^{\nu},\label{MaxeqRieam}
\end{eqnarray}
and more explicitly, in terms of the electric and magnetic components, we have
\begin{widetext}
\begin{eqnarray}
\dfrac{1}{c}\partial_{\mu}E_{j}\left(g^{0\mu}g^{j\nu}-g^{j\mu}g^{0\nu}\right)+\dfrac{1}{c}E_{j}\left[\partial_{\mu}\left(g^{0\mu}g^{j\nu}-g^{j\mu}g^{0\nu}\right)+\dfrac{1}{\sqrt{-g}}\partial_{\mu}(\sqrt{-g})\left(g^{0\mu}g^{j\nu}-g^{j\mu}g^{0\nu}\right)\right]  
    \nonumber \\
%
-\partial_{\mu}B^{k}g^{m\mu}g^{n\nu}\epsilon_{kmn}-B^{k}\epsilon_{kmn}\left[\partial_{\mu}(g^{m\mu}g^{n\nu})+\dfrac{1}{\sqrt{-g}}\partial_{\mu}(\sqrt{-g})(g^{m\mu}g^{n\nu}) \right]=\mu_{0}j^{\nu}.\label{Maxwellgeneral}
\end{eqnarray}
We now take the case where $\nu=0$, which gives the extended Gauss law in curved spacetime 
\begin{eqnarray}
\partial_{\mu}E_{j}\left(g^{0\mu}g^{j0}-g^{j\mu}g^{00}\right)+E_{j}\left[\partial_{\mu}\left(g^{0\mu}g^{j0}-g^{j\mu}g^{00}\right)+\dfrac{1}{\sqrt{-g}}\partial_{\mu}(\sqrt{-g})\left(g^{0\mu}g^{j0}-g^{j\mu}g^{00}\right)\right] 
   \nonumber \\ 
%
-\partial_{\mu}B^{k}cg^{m\mu}g^{n0}\epsilon_{kmn}-B^{k}c\epsilon_{kmn}\left[\partial_{\mu}(g^{m\mu}g^{n0})+\dfrac{1}{\sqrt{-g}}\partial_{\mu}(\sqrt{-g})(g^{m\mu}g^{n0}) \right]=\dfrac{\rho}{\varepsilon_{0}}.
   \label{GaussGenerall}
\end{eqnarray}
\end{widetext}
The same result can be obtained from Eq. (\ref{generalinhom}) via Eq. (\ref{contitutD}).
This equation generalizes Gauss' law and when compared with Maxwell's theory in Minkowski spacetime, it predicts new electromagnetic phenomena in the presence of sufficiently strong gravitational fields. In particular, in the absence of charge densities, even static magnetic fields can in principle induce an electric field. The presence of magnetic fields in Gauss' law disappear for vanishing off-diagonal time-space metric components. In fact, these correspond to the components of the gravitomagnetic potential defined in the weak field (linear) approximation of gravity (see \citep{GravitomagnetismMashhoon:2003ax}). Therefore, there is an electromagnetic coupling via gravitomagnetic effects. In the absence of such terms, $g^{0j}=g^{j0}=0$, we get 
\begin{eqnarray}
-\left(g^{jk}g^{00}\right)\partial_{k}E_{j}-E_{j}\Bigg[\partial_{k}\left(g^{jk}g^{00}\right)
    \nonumber \\
+\dfrac{1}{\sqrt{-g}}\partial_{k}(\sqrt{-g})g^{jk}g^{00}\Bigg]=\dfrac{\rho}{\varepsilon_{0}}, 
\end{eqnarray}
and for a diagonal metric, Gauss' law can be recast into the form
\begin{equation}
-g^{kk}g^{00}\partial_{k}E_{k}+E_{k}\gamma^{k}(\bold{x})=\dfrac{\rho}{\varepsilon_{0}},\label{gausssimple}
\end{equation}
with
\begin{equation}
\gamma^{k}(\bold{x})\equiv\ -\left( g^{kk}g^{00}\dfrac{1}{\sqrt{-g}}\partial_{k}(\sqrt{-g})+\partial_{k}(g^{kk}g^{00})\right),\label{funcgauss}
\end{equation}
and where no contraction is assumed in Eq. (\ref{funcgauss}). From Eq. (\ref{gausssimple}) we see that for Minkowski spacetime the usual Gauss law is clearly recovered. More interesting is the fact that for geometries where the functions $\gamma^{k}(\bold{x})$ are non-vanishing, the electric field in vacuum will necessarily be non-uniform, i.e., spacetime curvature will introduce an inevitable spatial variability in the electric field. In fact, suppose that in a given coordinate system the electric field has a single component $E_{j}$, then in vacuum we have $\gamma^{j}\neq 0$, which implies $\partial_{j}E_{j}\neq 0$.
%
%

We now consider the case where $\nu=i=1,2,3$ in Eq. (\ref{MaxeqRieam}) which corresponds to the extended or generalized Maxwell-Amp\`{e}re law in curved spacetime. We get
\begin{widetext}
\begin{eqnarray}
\dfrac{1}{c}\partial_{\mu}E_{j}\left(g^{0\mu}g^{ji}-g^{j\mu}g^{0i}\right)+\dfrac{1}{c}E_{j}\left[\partial_{\mu}\left(g^{0\mu}g^{ji}-g^{j\mu}g^{0i}\right)+\dfrac{1}{\sqrt{-g}}\partial_{\mu}(\sqrt{-g})\left(g^{0\mu}g^{ji}-g^{j\mu}g^{0i}\right)\right] 
    \nonumber  \\
%
-\partial_{\mu}B_{mn}g^{m\mu}g^{ni}-B_{mn}\left[\partial_{\mu}(g^{m\mu}g^{ni})+\dfrac{1}{\sqrt{-g}}\partial_{\mu}(\sqrt{-g})(g^{m\mu}g^{ni}) \right]=\mu_{0}j^{i},  \label{MaxAmpereGenerall}
\end{eqnarray}
and for the case where $g^{0j}=g^{j0}=0$, we obtain
\begin{eqnarray}
\dfrac{1}{c^{2}}\partial_{t}E_{j}\left(g^{00}g^{ji}\right)-\partial_{j}B_{mn}g^{mj}g^{ni}+\dfrac{1}{c^{2}}E_{j}\left[\partial_{t}\left(g^{00}g^{ji}\right)+\dfrac{1}{\sqrt{-g}}\partial_{t}(\sqrt{-g})\left(g^{00}g^{ji}\right)\right]  
    \nonumber  \\
%
-B_{mn}\left[\partial_{j}(g^{mj}g^{ni})+\dfrac{1}{\sqrt{-g}}\partial_{j}(\sqrt{-g})(g^{mj}g^{ni}) \right]=\mu_{0}j^{i}.
\end{eqnarray}
Finally the case of a diagonal metric becomes
\begin{eqnarray}
\dfrac{1}{c^{2}}\partial_{t}E_{i}\left(g^{00}g^{ii}\right)-\partial_{j}B_{ji}g^{jj}g^{ii}+\dfrac{1}{c^{2}}E_{i}\left[\partial_{t}\left(g^{00}g^{ii}\right)+\dfrac{1}{\sqrt{-g}}\partial_{t}(\sqrt{-g})\left(g^{00}g^{ii}\right)\right]  
    \nonumber  \\
%
-B_{ji}\left[\partial_{j}(g^{jj}g^{ii})+\dfrac{1}{\sqrt{-g}}\partial_{j}(\sqrt{-g})(g^{jj}g^{ii}) \right]=\mu_{0}j^{i},
\end{eqnarray}
\end{widetext}
which may be written in the following form
\begin{eqnarray}
\epsilon_{ijk}g^{ii}g^{jj}\partial_{j}B^{k}&+&\dfrac{1}{c^{2}}g^{00}g^{ii}\partial_{t}E_{i}
   \nonumber  \\
&+&\epsilon_{ijk}\sigma^{ji
i}B^{k}+\dfrac{1}{c}E_{i}\xi^{ii}=\mu_{0}j^{i},\label{MaxAmp2}
\end{eqnarray}
where the Einstein summation convention is applied in this expression only for $j$ and $k$ while the index $i$ is fixed by the right hand side and
\begin{eqnarray}
\sigma^{jii}(\bold{x}) \equiv  g^{jj}g^{ii}\dfrac{1}{\sqrt{-g}}\partial_{j}(\sqrt{-g})+\partial_{j}(g^{jj}g^{ii}),\\
\xi^{ii}(\bold{x}) \equiv  g^{00}g^{ii}\dfrac{1}{c}\dfrac{1}{\sqrt{-g}}\partial_{t}(\sqrt{-g})+\dfrac{1}{c}\partial_{t}(g^{00}g^{ii}).\label{sigmaximaxamp}
\end{eqnarray}
Again no contraction is assumed in the above expression for $\sigma^{jii}$.

While in Gauss' law the electromagnetic coupling disappears for a diagonal metric, in the Maxwell-Amp\`{e}re law this coupling is always present. In fact, the two terms containing the electric field are intertwined. Although for stationary geometries the term proportional to the electric field vanishes ($\xi^{ii}=0$), while the usual term with the time derivative can still be present (for that to happen a time varying charge density in Gauss' law is sufficient), the opposite is not true. That is, for dynamical time varying geometries, these two terms come together since a non-stationary spacetime will necessarily induce a time varying electric field via Gauss' law. Accordingly, gravitational waves are expected to produce a direct effect in magnetic fields.
As in the case of Gauss' law, new predictions emerge due to curved spacetime geometry. For vanishing currents the presence of an electric field  can be a source of  magnetic fields, with an extra contribution to Maxwell's displacement current induced by spacetime dynamics, coming from the functions $\xi^{ii}$. These functions vanish for a stationary spacetime but might have an important contribution for strongly varying gravitational waves (high frequencies). Correspondingly, the Maxwell-Amp\`{e}re law can be rewritten in the following form
\begin{equation}
\epsilon_{ijk}\partial_{j}\bar{B}^{iijjk}=\mu_{0}(\jmath^{i}+\jmath^{i}_{D}),
\end{equation}
where 
\begin{equation}
\jmath^{i}\equiv \sqrt{-g}j^{i},\qquad
\jmath^{i}_{D}\equiv -\varepsilon_{0}\sqrt{-g}\left(g^{00}g^{ii}\partial_{t}E_{i}+cE_{i}\xi^{ii}\right), 
\end{equation}
and
\begin{equation}
\bar{B}^{iijjk}\equiv g^{ii}g^{jj}\sqrt{-g}B^{k}. 
\end{equation}

\subsubsection{Homogeneous equations}

Explicitly, the homogeneous equations are given by
\begin{eqnarray}
\nabla_{\alpha}F_{\beta\gamma}&+&\nabla_{\beta}F_{\gamma\alpha}+\nabla_{\gamma}F_{\alpha\beta}=\partial_{\alpha}F_{\beta\gamma}+\partial_{\beta}F_{\gamma\alpha}+\partial_{\gamma}F_{\alpha\beta}
     \nonumber   \\
&&-F_{\beta\lambda}\Gamma^{\lambda}_{[\gamma\alpha]}-F_{\alpha\lambda}\Gamma^{\lambda}_{[\beta\gamma]}-F_{\gamma\lambda}\Gamma^{\lambda}_{[\alpha\beta]}=0.
\end{eqnarray}
Therefore, for torsionless manifolds, as is the case of a spacetime with Riemann geometry, these equations give the usual Faraday law ($\partial_{t}B^{i}=-\epsilon^{ijk}\partial_{j}E_{k}$) and the Gauss law for magnetism ($\partial_{j}B^{j}=0$), which are unaffected by spacetime geometry as long as spacetime torsion is zero.

\subsubsection{Equations for the potential and electromagnetic waves}

Let us review the inhomogeneous equations in terms of the electromagnetic potential. We start with the following expression
\begin{equation}
\nabla_{\mu}F^{\mu\nu}=\nabla_{\mu}\nabla^{\mu}A^{\nu}-\left( \left[\nabla_{\mu},\nabla_{\lambda}\right]A^{\mu}+\nabla_{\lambda}\nabla_{\mu}A^{\mu}\right )g^{\lambda\nu},   
\end{equation}
and since $\left[\nabla_{\mu},\nabla_{\lambda}\right]A^{\alpha}=R^{\alpha}_{\,\varepsilon\mu\lambda}A^{\varepsilon}$, where $R^{\alpha}_{\,\varepsilon\mu\lambda}$ and $R^{\mu}_{\,\varepsilon\mu\lambda}\equiv R_{\varepsilon\lambda}$ are the components of the Riemann and Ricci tensors respectively, we get the following Maxwell equation
\begin{equation}
\nabla_{\mu}\nabla^{\mu}A^{\nu}-g^{\lambda\nu}R_{\varepsilon\lambda}A^{\varepsilon}-\nabla^{\nu}\left( \nabla_{\mu}A^{\mu}\right)=\mu_{0}j^{\nu}.  
\end{equation}
For potentials satisfying the Lorentz condition (in curved spacetime)
\begin{equation}
\nabla_{\mu}A^{\mu}=\dfrac{1}{\sqrt{-g}}\partial_{\mu}\left( \sqrt{-g}A^{\mu}\right)=0,
\end{equation}
we get
\begin{equation}
\nabla_{\mu}\nabla^{\mu}A^{\nu}-g^{\lambda\nu}R_{\varepsilon\lambda}A^{\varepsilon}=\mu_{0}j^{\nu}.   
\end{equation}
Using the expression for the (generalized) Laplacian in pseudo-Riemann manifolds,
\begin{equation}
\nabla_{\mu}\nabla^{\mu}\psi=
\dfrac{1}{\sqrt{-g}}\partial_{\mu}\left( \sqrt{-g}g^{\mu\lambda}\partial_{\lambda}\psi\right), 
\end{equation}
we arrive at
\begin{equation}
\partial_{\mu}\partial^{\mu}A^{\nu}+
\dfrac{1}{\sqrt{-g}}\partial_{\mu}\left( \sqrt{-g}g^{\mu\lambda}\right)\partial_{\lambda}A^{\nu}-g^{\lambda\nu}R_{\varepsilon\lambda}A^{\varepsilon}=\mu_{0}j^{\nu},\label{Proca}   
\end{equation}
which in vacuum is a generalized Proca-like equation with variable (spacetime dependent) effective mass induced by the curved geometry. The second term in Eq. (\ref{Proca}) can also be written in terms of the Levi-Civitta connection, through the formula $g^{\alpha\beta}\Gamma^{\lambda}_{\,\alpha\beta}=-\dfrac{1}{\sqrt{-g}}\partial_{\alpha}\left( \sqrt{-g}g^{\alpha\lambda}\right)$, valid in pseudo-Riemann geometry. In usual Proca-like wave equations there is no such term dependent on the first derivative of the (massive) vector field. Similiar terms appear for wave phenomena with longitudinal modes. For a diagonal metric in vacuum we get,
\begin{equation}
\partial_{\mu}\partial^{\mu}A^{\nu}+
\dfrac{1}{\sqrt{-g}}\partial_{\mu}\left( \sqrt{-g}g^{\mu\mu}\right)\partial_{\mu}A^{\nu}-g^{\nu\nu}R_{\varepsilon\nu}A^{\varepsilon}=0, 
\end{equation}
with no contraction assumed in $\nu$. In general, and contrary to electromagnetism in Minkowski spacetime, the equations for the components of the electromagnetic 4-potential are coupled even in the Lorentz gauge. Notice also that for Ricci-flat spacetime, the term containing the Ricci tensor vanishes. Naturally, the vaccum solutions of GR are examples of such cases. New electromagnetic phenomena are expected to be measurable, for gravitational fields where the geometry dependent terms in Eq. (\ref{Proca}) are significant.

We now consider the appropriate gauge-invariant expressions describing electromagnetic waves. Using Eq. (\ref{Maxeqriem1}), the vacuum field equations are given by
\begin{eqnarray}
\nabla_{\mu}F^{\mu\nu}&=&\partial_{\mu}F^{\mu\nu}+\dfrac{1}{\sqrt{-g}}\partial_{\mu}(\sqrt{-g})F^{\mu\nu}=0,
    \\
\nabla_{[\alpha}F_{\beta\gamma]} &=& 0,   
\end{eqnarray}
which are gauge invariant.
If we consider the Maxwell-Amp\`{e}re law in vacuum ($\nu=i$) and derive it with respect to time and then using the Faraday law and also Gauss' law in vacuum ($\nu=0$), one arrives at the following generalized (gauge invariant) wave equation for a diagonal metric
\begin{eqnarray}
&&g^{ii}\left( \dfrac{g^{00}}{c^{2}}\partial^{2}_{tt}E_{i}+\partial^{k}\partial_{k}E_{i}    \right)=a^{ii}\partial_{t}E_{i}+\tilde{b}^{kii}\partial_{k}E_{i}
    \nonumber \\
&&+\bar{b}^{kii}\partial_{i}E_{k}+\beta^{kki}\partial_{k}E_{k}+c^{ii}E_{i}+c^{ki}_{2}E_{k}
    \nonumber \\
&&+m^{kkii}\partial_{k}B_{ki}+n^{kii}B_{ki}. \label{waveelectric} 
\end{eqnarray}
Here there is no contraction on the index $i$ (only on $k$), while the coefficients which depend on spacetime geometry are given by the following expressions (without applying the Einstein summation convention)
\begin{eqnarray}
a^{ii}&\equiv &-\dfrac{1}{c}\left[2\partial_{t}(g^{00}g^{ii})+\dfrac{g^{00}g^{ii}}{\sqrt{-g}}\partial_{t}(\sqrt{-g}) \right],
\nonumber
\end{eqnarray}
\begin{eqnarray}
 \tilde{b}^{kii}&\equiv & -\left[ \partial_{k}(g^{kk}g^{ii})+\dfrac{g^{kk}g^{ii}}{\sqrt{-g}}\partial_{k}(\sqrt{-g})\right],
\nonumber
\end{eqnarray}
\begin{equation}
\bar{b}^{kii}\equiv g^{ii}g_{00}\gamma^{k}+\left[ \partial_{k}(g^{kk}g^{ii})+\dfrac{g^{kk}g^{ii}}{\sqrt{-g}}\partial_{k}(\sqrt{-g})\right],
\nonumber
\end{equation}
\begin{equation}
 \beta^{kki}\equiv -g^{ii}\partial_{i}g^{kk},
\nonumber
\end{equation}
\begin{equation}
c^{ii}\equiv -\left[\dfrac{1}{c}\partial^{2}_{tt}(g^{00}g^{ii})+\dfrac{1}{c}\partial_{t}\left[ g^{00}g^{ii}\left( \dfrac{\partial_{t}(\sqrt{-g})}{\sqrt{-g}}\right)\right]\right],
\nonumber
\end{equation}
\begin{equation}
 c^{ki}_{2}\equiv g^{ii}\partial_{i}(g_{00}\gamma^{k}), \qquad  m^{kkii}\equiv \partial_{t}(g^{ii}g^{kk}), 
\nonumber
\end{equation}
\begin{equation}
  n^{kii}\equiv\partial_{k}\partial_{t}(g^{ii}g^{kk})+ \partial_{t}\left( \dfrac{\partial_{k}(\sqrt{-g})}{\sqrt{-g}}g^{ii}g^{kk}\right), 
\nonumber
\end{equation}
and $\gamma^{k}$ is given by Eq. (\ref{funcgauss}).
We therefore get coupled electromagnetic waves where the coupling is induced by the non-flat and dynamical character of spacetime geometry. The right-hand-side (rhs) of the wave equation vanishes in flat spacetime. Notice that the magnetic terms are present for time varying geometries and therefore, one expects gravitational waves to couple the electric and magnetic wave dynamics with measurable consequences. This coupling is likely to affect polarization and interference patterns, but requires further research.

A similar expression can be obtained deriving the Faraday law with respect to time and then using the Maxwell-Amp\`{e}re and (magnetic) Gauss laws. Nevertheless, while Maxwell's equations in vacuum, and in Minkowski spacetime, have a complete symmetry regarding the electric and magnetic fields, here since $d\bold{F}=0$ is independent from the metric structure while $d\bold{\star F}=0$ is not, the fundamental difference results since $\partial_{m}B^{m}=0$, while in general $\partial_{k}E^{k}\neq 0$.  

Assuming a diagonal metric we see that such a wave equation for the magnetic field components is also coupled to the electric field, for non-stationary metric,
\begin{eqnarray}
\dfrac{g^{00}}{c^{2}}\partial^{2}_{tt}B^{j}+\partial^{k}\partial_{k}B^{j}&=&\tilde{\eta}^{jm}_{1}E_{m}+\tilde{\eta}_{2}\partial_{t}B^{j}+\tilde{\alpha}^{k}\partial_{k}B^{j}\nonumber \\
&+&\tilde{\delta}^{jj}_{k}\partial_{j}B^{k}+\tilde{c}B^{j}+\tilde{c}^{j}_{k}B^{k},\label{wave magnetic}
\end{eqnarray}
with coefficients
\begin{equation}
\tilde{\eta}^{jm}_{1}=g^{00}\epsilon^{jkm}\partial_{k}(g_{00}g_{mm}\xi^{mm}),\quad \tilde{\eta}_{2}=-g_{nn}\xi^{nn},
 \nonumber 
\end{equation}
\begin{equation}
 \tilde{\alpha}^{k}=-\left( g^{00}\partial_{k}(g_{00}g^{kk})+\sigma^{mmk}g_{mm}\right),
  \nonumber
\end{equation}
\begin{equation}
\tilde{\delta}^{jj}_{k}= g^{00}\partial_{k}(g_{00}g^{jj}),\quad  \tilde{c}=- g^{00}\partial_{k}(\sigma^{mmk}g_{00}g_{mm}),
   \nonumber
\end{equation}
\begin{equation}
\tilde{c}^{j}_{k}= g^{00}\partial_{k}(\sigma^{mmj}g_{00}g_{mm}),
   \nonumber
\end{equation}
where $\xi^{mm}$ is given by Eq. (\ref{sigmaximaxamp}) and the Einstein summation convention is only applied in $\tilde{\eta}_{2}$, $\tilde{c}$ and $\tilde{c}^{j}_{k}$.

For a stationary (or static) spacetime, both wave equations become decoupled since $a^{ii}=c^{ii}=m^{kkii}=n^{kii}=0$ in (\ref{waveelectric}) and $\tilde{\eta}^{jm}_{1}= \tilde{\eta}_{2}=0$ in Eq. (\ref{wave magnetic}), leading to
\begin{eqnarray}
 g^{ii}\left( \dfrac{g^{00}}{c^{2}}\partial^{2}_{tt}E_{i}+\partial^{k}\partial_{k}E_{i}    \right)&=&\tilde{b}^{kii}\partial_{k}E_{i}+\bar{b}^{kii}\partial_{i}E_{k}
\nonumber \\
&+& \beta^{kki}\partial_{k}E_{k}+c^{ki}_{2}E_{k},\label{wave stationary space}
\end{eqnarray}
\begin{equation}
 \dfrac{g^{00}}{c^{2}}\partial^{2}_{tt}B^{j}+\partial^{k}\partial_{k}B^{j}=\tilde{\alpha}^{k}\partial_{k}B^{j}+\tilde{\delta}^{jj}_{k}\partial_{j}B^{k}+\tilde{c}B^{j}+\tilde{c}^{j}_{k}B^{k}.\label{wave magnetic stationary}
\end{equation}
We can see from these equations that for stationary geometries the wave equations for the electric and magnetic fields are decoupled, but in each case there is a coupling between different components, which suggests polarization effects. By careful comparison between these two equations, or between Eqs. (\ref{waveelectric}) and (\ref{wave magnetic}), we also see that the symmetry is not complete which, as previously mentioned, is essentially rooted in the fact that generally the magnetic field is divergenceless, contrary to the electric field.

As an example, let us consider the case where the electric field has a single component $E_{j}$ (where $j$ is fixed). This could happen, for example, for a linearly polarized wave or for a general polarization where the symmetries allow a curvilinear coordinate system to follow the field, so that in such system it would have a single component, for example, $\bold{E}=E_{\varphi}(z)$. In such case, from (\ref{wave stationary space}) we get in Fourrier space
\begin{eqnarray}
&& g^{jj}\left( \dfrac{-w^{2}}{c^{2}}g^{00}\tilde{E}_{j}+\partial^{k}\partial_{k}\tilde{E}_{j}\right)=
\nonumber \\
&&(\tilde{b}^{jjj}+\bar{b}^{jjj}+\beta^{jjj})\partial_{j}\tilde{E}_{j}+c^{jj}_{2}\tilde{E}_{j},
\end{eqnarray}
therefore
\begin{eqnarray}
 g^{jj}\partial^{k}\partial_{k}\tilde{E}_{j} &-& (\tilde{b}^{jjj}+\bar{b}^{jjj}+\beta^{jjj})\partial_{j}\tilde{E}_{j}
\nonumber \\
&-& \left(g^{jj} \dfrac{w^{2}}{c^{2}}g^{00}+c^{jj}_{2}\right)\tilde{E}_{j}=0,
\end{eqnarray}
for each mode, where $\tilde{E}_{j}=\tilde{E}_{j}(x^{1},x^{2},x^{3},w)$ is the Fourier transform of $E_{j}$. The above equation resembles a generalized time independent Schrodinger equation or a generalized Laplace equation with corrections induced by spacetime geometry. Recall that here $j$ is fixed, with no contraction, and only $k$ is a summation index.
For a wave travelling along the $x$ axis, we get
\begin{eqnarray} 
g^{jj}\partial^{2}_{xx}\tilde{E}_{j}&-&(\tilde{b}^{jjj}+\bar{b}^{jjj}+\beta^{jjj})\partial_{j}\tilde{E}_{j}
\nonumber \\
&-& \left(g^{jj} \dfrac{w^{2}}{c^{2}}g^{00}+c^{jj}_{2}\right)\tilde{E}_{j}=0.
\end{eqnarray}
So the equations for the cases of transversal or (hypothetical) longitudinal modes become
\begin{equation}
g^{jj}\partial^{2}_{xx}\tilde{E}_{j}-\left(g^{jj} \dfrac{w^{2}}{c^{2}}g^{00}+c^{jj}_{2}\right)\tilde{E}_{j}=0,
\end{equation}
\begin{eqnarray}
 g^{xx}\partial^{2}_{xx}\tilde{E}_{x}&-&(\tilde{b}^{xxx}+\bar{b}^{xxx}+\beta^{xxx})\partial_{x} \tilde{E}_{x}
   \nonumber \\
&-& \left(g^{xx} \dfrac{w^{2}}{c^{2}}g^{00}+c^{xx}_{2}\right)\tilde{E}_{x}=0,
\end{eqnarray}
respectively (where $j=y,z$). By introducing the variables $Y_{1}\equiv \tilde{E}$, $Y_{2}
\equiv \partial_{x}\tilde{E}$, these equations can be put in the form of a linear 
non-autonomous dynamical system (with one degree of freedom) according to
\begin{equation}
\partial_{x}\vec{Y}=\bold{A}\vec{Y},\label{dynamicalsystem}
\end{equation}
where in the first case, 
\begin{equation}
A_{11}=A_{22}=0, \quad A_{12}=1, \quad A_{21}=\dfrac{w^{2}}{c^{2}}g^{00}+c^{jj}_{2}g_{jj},
  \nonumber 
\end{equation}
%
with ($j=y,z$), while in the second system 
\begin{eqnarray}
A_{11}=0, \quad A_{22}=(\tilde{b}^{xxx}+\bar{b}^{xxx}+\beta^{xxx})g_{xx},
   \nonumber \\
A_{12}=1, \quad A_{21}=\dfrac{w^{2}}{c^{2}}g^{00},
+c^{xx}_{2}g_{xx}.
  \nonumber
\end{eqnarray}
This leads to the following expressions for the eigenvalues
\begin{equation}
\lambda=\pm\sqrt{A_{21}}, \qquad \lambda=\dfrac{A_{22}\pm\sqrt{A_{22}^{2}+4A_{21}}}{2},
\end{equation}
respectively.

The system in Eq. (\ref{dynamicalsystem}) is non-autonomous, in general, and becomes autonomous if the direction of propagation of the waves has a correspondence with a spacetime isometry (as there is a killing vector along that direction). In such a scenario, the theory of dynamical systems can easily provide the qualitative behaviour of the dynamics. For transversal modes, depending on the background spacetime, we will get real eigenvalues with opposite signs or purely imaginary eigenvalues. Therefore, the dynamics is that of a (global) phase space with a saddle fixed point at (0,0), which corresponds to a decaying $f$ with no oscillations, or a purely periodic behaviour around a center fixed point at (0,0). For the longitudinal modes the system predicts a richer behaviour around the (0,0) fixed point, depending on the geometry, with physical acceptable solutions corresponding to purely oscillatory ($A_{22}=0$, $A_{22}^{2}+4A_{21}<0$), decaying and oscillatory decaying behaviour. 

We therefore expect new phenomena associated with electromagnetic waves in the presence of a background gravitational field. In particular, the wave equations presented here can be applied to different geometries such as the spacetime outside a spherical gravitating mass (Schwarzschild solution of GR) and the corresponding linear (weak field) limit, for the time-independent case and gravitational waves with (+) polarization, for the time dependent scenario. 
It is natural to expect an effect in the polarization of electromagnetic waves induced by the curvature of (pseudo) Riemann spacetime (se also \cite{Hacyan:2015kra}). We see that in Eqs. (\ref{wave stationary space}) and (\ref{wave magnetic stationary}), that due to the time-independent character of spacetime geometry, the wave equation has no electromagnetic coupling, nevertheless, the last three terms in the rhs of Eq. (\ref{wave stationary space}) and the second and fourth terms in the rhs of Eq. (\ref{wave magnetic stationary}) imply a coupling between the dynamics of the different field components. As previously mentioned, this fact suggests polarization effects induced by the spacetime geometry.

What about other effects? Are there any predicted longitudinal modes besides the transversal modes characterizing electromagnetic waves in flat Minkowski spacetime? These longitudinal modes in vector ($a^{k}$) wave equations appear whenever $\partial_{k}a^{k}\neq0$, and in fact, in usual electromagnetism such terms would appear if the electric and magnetic fields in vacuum weren't divergenceless. Accordingly, in curved spacetime, $\partial_{k}E^{k}\neq0$, which manifests in the terms containing the first derivatives of the electric field in Eq. (\ref{wave stationary space}). These longitudinal modes seem to be a prediction of electromagnetic wave dynamics in curved spacetime.   
In turns out that if harmonic waves with constant amplitudes are solutions to the wave equations then, in general, electromagnetic waves are no longer purely  transversal. This can be proven by following the method which is often used to show the transversal character of waves in flat spacetime. 

Consider a harmonic wave given by
\begin{equation}
E_{i}=E_{0i}e^{-ik^{\mu}x_{\mu}},
\end{equation}
We then have,
\begin{eqnarray}
k^{\mu}x_{\mu}&=&g_{00}wt + g_{0m}\left(k^{0}x^{m} + k^{m}ct \right)+g_{rs}k^{r}x^{s},
\\
 k^{\mu}k_{\mu}&=&0, 
\end{eqnarray}
and for a general curved spacetime we will allow the components of the wave four-vector to be spacetime dependent [$k^{\mu}=k^{\mu}(x^{m},t)$]. This can be understood by considering any two spacetime points, $(x_{1}^{m},t_{1})$ and $(x_{2}^{m},t_{2})$, along the light path such that the sine and cosine functions are equal. This periodicity implies that 
\begin{eqnarray}
&& w=\dfrac{2\pi}{\tau},\quad \sqrt{k^{j}k_{j}}=\dfrac{2\pi}{\lambda},\\
&& \tau=\int_{t_{1}}^{t_{2}}\sqrt{g_{00}}dt,\quad \lambda=\int_{X_{1}}^{X_{2}}\sqrt{g_{kk}}dx^{k},
\end{eqnarray}
and therefore, in general, the period and the wavelength will depend on the spacetime point at which they are computed. 
Consequently
\begin{eqnarray}
\partial_{t}(k^{\mu}x_{\mu})&=&(\partial_{t}g_{00})wt
+g_{00}w+\partial_{t}(g_{0m})\left(k^{0}x^{m}+k^{m}ct \right)
    \nonumber  \\
&&+g_{0m}k^{m}c+\partial_{t}(g_{rs})k^{r}x^{s}+(\partial_{t}k^{\mu})x_{\mu}, 
\end{eqnarray}
\begin{eqnarray}
\partial_{j}(k^{\mu}x_{\mu})&=&(\partial_{j}g_{00})wt
+\partial_{j}(g_{0m})\left(k^{0}x^{m}+k^{m}ct \right)+g_{0j}k^{0}
   \nonumber \\
&&+\partial_{j}(g_{rs})k^{r}x^{s}+g_{rj}k^{r}+(\partial_{j}k^{\mu})x_{\mu}. 
\end{eqnarray}
For simplicity and without a significant loss of generality, let us consider a diagonal metric. In this case, we have
\begin{eqnarray}
\partial_{t}(k^{\mu}x_{\mu})=(\partial_{t}g_{00})wt
+g_{00}w+(\partial_{t}g_{rr})k^{r}x^{r}
    \nonumber \\
+g_{00}t\partial_{t}w+g_{rr}x^{r}\partial_{t}k^{r}, 
\end{eqnarray}
\begin{eqnarray}
\partial_{j}(k^{\mu}x_{\mu})=(\partial_{j}g_{00})wt
+(\partial_{j}g_{rr})k^{r}x^{r}+g_{jj}k^{j}
   \nonumber \\
+g_{00}t\partial_{j}w+g_{rr}x^{r}\partial_{j}k^{r}. 
\end{eqnarray}
Now, Gauss' law (\ref{gausssimple}) gives
\begin{equation}
g^{jj}\partial_{j}E_{j}=g_{00}\gamma^{j}E_{j}.
\end{equation}
On the other hand, for the harmonic electromagnetic wave, we have
\begin{equation}
\partial_{j}E_{j}=-i\partial_{j}(k^{\mu}x_{\mu})E_{j},
\end{equation}
and therefore,
\begin{eqnarray}
g_{00}\gamma^{j}E_{j}&=&-i \big[ (\partial_{j}g_{00})wt
+(\partial_{j}g_{rr})k^{r}x^{r}+g_{jj}k^{j}
    \nonumber  \\
&&+g_{00}t\partial_{j}w+g_{rr}x^{r}\partial_{j}k^{r}\big] g^{jj}E_{j}.
\end{eqnarray}
This implies the following
\begin{eqnarray}
k^{j}E_{j}&=&-ig_{00}\gamma^{j}E_{j}-\big[(\partial_{j}g_{00})wt
+(\partial_{j}g_{rr})k^{r}x^{r}
   \nonumber  \\
&&+g_{00}t\partial_{j}w+g_{rr}x^{r}\partial_{j}k^{r}\big] g^{jj}E_{j},  
\end{eqnarray}
and consequently
\begin{eqnarray}
k^{j}E_{0j}&=&-\Big[ ig_{00}\gamma^{j}+g^{jj}\big((\partial_{j}g_{00})wt
+(\partial_{j}g_{rr})k^{r}x^{r}
    \nonumber  \\
&&+g_{00}t\partial_{j}w+g_{rr}x^{r}\partial_{j}k^{r}\big) \Big]E_{0j}.  
\end{eqnarray}
Since the terms on the rhs are in general different from zero, these expressions imply that
\begin{equation}
Re(k^{j}E_{j})=f(t,x^{m})\neq 0.
\end{equation}
For example, consider a harmonic wave propagating in the $z$ direction ($k_{x}=k_{y}=0$). Maxwell's equations merely imply that
\begin{widetext}
\begin{eqnarray}
k^{z}E_{0z}=-\dfrac{\left( ig_{00}\gamma^{x}+g^{xx}\left( \left(\partial_{x}g_{00}\right) wt+\partial_{x}(g_{zz})k^{z}z+g_{00}t\partial_{x}w+g_{zz}z\partial_{x}k^{z}\right)\right) E_{0x}}{\left(1+ (\partial_{z}g_{zz})zg^{zz}\right)}
    \nonumber  \\
%
-\dfrac{\left( ig_{00}\gamma^{y}+g^{yy}\left( \left(\partial_{y}g_{00}\right) wt+(\partial_{y}g_{zz})k^{z}z+g_{00}t\partial_{y}w+g_{zz}z\partial_{y}k^{z}\right)\right) E_{0y}}{\left(1+ (\partial_{z}g_{zz})zg^{zz}\right)}
    \nonumber  \\
%
-\dfrac{\left( ig_{00}\gamma^{z}+g^{zz}\left( \left(\partial_{z}g_{00}\right) wt+g_{00}t\partial_{z}w+g_{zz}z\partial_{z}k^{z}\right) \right)E_{0z}}{\left(1+(\partial_{z}g_{zz})zg^{zz}\right)},\label{kzez}  
\end{eqnarray}
\end{widetext}
which, in general, is non-zero. Thus, one could in principle have a harmonic wave with a non-vanishing $E_{0z}$ (longitudinal component) without transgressing the mathematical-physics of Maxwell's theory in Riemann spacetime.

If the frequency $w$ is known, the real part of the previous equation provides a differential equation for $k^{z}$ providing an important application for rigorous cosmological redshift calculations which considers electrodynamics in curved spacetime,
\begin{equation}
\chi^{j}\partial_{j}\bar{k}^{z}+\chi\bar{k}^{z}+\Lambda=0
\end{equation}
where $\bar{k}_{z}\equiv Re (k^{z})$ and the factors $\chi^{j},\chi,\Lambda$ depend on the spacetime metric, the frequency and the amplitudes are taken from Eq. (\ref{kzez}).

In general, the harmonic wave is not a plane transversal wave ($k^{j}E_{j}\neq 0$) and one predicts longitudinal modes very naturally. Therefore not only the deflection of light and gravitational redshift but also  other effects are expected to result from the interaction of light and gravity (curved spacetime) such as polarization effects and the appearance of longitudinal modes.

For the sake of completeness, we present the expression for electromagnetic waves in the more general case without restricting to a diagonal metric. This can then be applied to axially symmetric spacetimes (as for the geometry outside a rotating mass in the weak field and slow rotation regime), as well to gravitational waves. In the first case, we usually have $g^{j0}\neq 0$, $g^{jk}=0$ for $j\neq k$, while in the second case  $g^{j0}=0$, $g^{jk}\neq 0$ for $j\neq k$.

The wave equation is obtained by deriving Eq. (\ref{MaxAmpereGenerall}) with respect to time and using the Faraday law, and one obtains
\begin{widetext}
\begin{eqnarray}
\dfrac{\bar{\Sigma}^{0ji}}{c^{2}}\partial^{2}_{tt}E_{j}-g^{mk}g^{ni}(\partial_{k}\partial_{n}E_{m}-\partial_{k}\partial_{m}E_{n})-(\partial_{n}E_{m}-\partial_{m}E_{n})\left[ \dfrac{1}{c}\partial_{t}(g^{m0}g^{ni})+\bar{\sigma}^{mni}\right]
+(\partial_{t}\bar{\Sigma}^{kji})\partial_{k}E_{j}
     \nonumber  \\
+\bar{\Sigma}^{kji}\partial_{t}\partial_{k}E_{j}
%
-\dfrac{g^{m0}g^{ni}}{c}\partial_{t}(\partial_{n}E_{m}-\partial_{m}E_{n})+\partial_{t}E_{j}\left(\dfrac{1}{c^{2}}\partial_{t}\bar{\Sigma}^{0ji}+\dfrac{1}{c}\bar{\xi}^{ji} \right)
+\dfrac{(\partial_{t}\bar{\xi}^{ji})}{c}E_{j}
    \nonumber  \\
-\epsilon_{mnj}\left[ \partial_{t}(g^{mk}g^{ni})\partial_{k}B^{j}+(\partial_{t}\bar{\sigma}^{mni})B^{j}\right] =0, \label{wavenondiagonal} 
\end{eqnarray}
\end{widetext}
where
\begin{equation}
\bar{\Sigma}^{0ji}=g^{00}g^{ji}-g^{j0}g^{0i}, \quad \bar{\Sigma}^{kji}=g^{0k}g^{ji}-g^{jk}g^{0i}, \nonumber
\end{equation}
\begin{equation}
\bar{\xi}^{ji}=\partial_{\mu}(g^{0\mu}g^{ji}-g^{j\mu}g^{0i})+\dfrac{1}{\sqrt{-g}}\partial_{\mu}(\sqrt{-g})(g^{0\mu}g^{ji}-g^{j\mu}g^{0i}),\nonumber
\end{equation}
\begin{equation}
\bar{\sigma}^{mni}=\partial_{\mu}(g^{m\mu}g^{ni})+\dfrac{1}{\sqrt{-g}}\partial_{\mu}(\sqrt{-g})(g^{m\mu}g^{ni}). \nonumber
\end{equation}

As one can see, this wave equation, although linear is much more complicated and should include a richer set of electromagnetic wave phenomena. Note that in principle, the terms containing the first derivatives of the electric field can be developed using Gauss' law, which from Eq. (\ref{GaussGenerall}) can be written as follows
\begin{eqnarray}
\bar{\Sigma}^{kj0}\partial_{k}E_{j}&=&-\bar{\gamma}^{j}E_{j}
+g^{m0}g^{n0}c\partial_{t}B_{mn}
  \nonumber \\
&&+g^{mj}g^{n0}c\partial_{j}B_{mn}+\bar{\sigma}^{mn0}cB_{mn},
\end{eqnarray}
with $\bar{\Sigma}^{kj0}$, $\bar{\gamma}^{j}$ and $\bar{\sigma}^{mn0}$, obtained by replacing $i$ with $0$ in $\bar{\Sigma}^{kji}$, $\bar{\xi}^{ji}$ and $\bar{\sigma}^{mni}$, respectively.

\subsection{Charge conservation and the Lorentz force}

\paragraph{Charge conservation in curved spacetime:}
 We saw that charge conservation alone implies the inhomogeneous equations.  In the tensor formalism charge conservation requires the spacetime connection due to the co-variant derivative. Charge conservation in components have then an explicit dependence on the geometry. In fact, from $d\bold{J}=0$, we get
\begin{equation}
\nabla_{[\mu}J_{\alpha\beta\gamma]}=0 \quad  \Leftrightarrow \quad 
\nabla_{\nu} j^{\nu}=0,
\end{equation}
therefore, according to
\begin{equation}
\partial_{t}\rho+\partial_{k}j^{k}=-( \Gamma^{\mu}_{\,\mu 0}\rho c+ \Gamma^{\mu}_{\,\mu k}j^{k}),
\end{equation}
even if there is no electric current, a non-static spacetime will induce a time variability in the charge density
\begin{equation}
\partial_{t}\rho=- \Gamma^{\mu}_{\,\mu 0}\rho c=-\partial_{t}(\log (\sqrt{-g})\rho,
\end{equation}
and therefore
\begin{equation}
\rho(\bold{x})=K\dfrac{1}{\sqrt{g}},
\end{equation}
where $K$ is an integration constant. More specifically, if the integration is performed from $t_{0}$ to $t$ we get
\begin{equation}
\rho(\bold{x})=\rho_{0}\sqrt{\dfrac{g_{0}}{g(t)}},\label{chargedynamical}
\end{equation}
where $\rho_{0}$ is the initial charge density which might be a function of space coordinates for initially non-uniform charge distributions and $g_{0}$ is the determinant of the initial metric.
This time variability will not be due to the motion of charges in currents but rather to the deformation of spacetime geometry. Therefore, again we expect possible measurable effects of gravitational waves in electrodynamics. The passage of a gravitational wave with Weyl curvature will distort shapes while preserving volumes. It is natural to expect a measurable effect on initially static charge distributions. It also seems natural to expect that the electric field lines (initially static) generated by such charge distributions will also follow the deformations of spacetime, according to Gauss' law. 
\\

\paragraph{Lorentz force in curved spacetime:}
We now consider the motion of a charged test particle under the influence of electromagnetic fields in a curved background spacetime. The equation of motion is given by
\begin{equation}
m\nabla_{\bold{u}}\bold{u}=\bold{f},
\end{equation}
where $\nabla_{\bold{u}}\bold{u}$ is the covariant derivative of the particle's 4-velocity $\bold{u}$ with respect to itself and $\bold{f}$ is the electromagnetic force 4-vector. In the absence of electromagnetic fields, the particle follows the auto-parallel geodesics $\nabla_{\bold{u}}\bold{u}=0$ which, as previously mentioned, coincide with the extremal geodesics in the spacetime of GR (the same is not true if torsion is present with an extra term coming from the symmetric part of the contorsion tensor). The above equation reads
\begin{equation}
m\left( \dot{u}^{\lambda}+\Gamma^{\lambda}_{\;\alpha\beta}u^{\alpha}u^{\beta}\right)=qF^{\lambda}_{\;\nu}u^{\nu}=qF_{\mu\nu}g^{\mu\lambda}u^{\nu}.\label{Lorentzforce}
\end{equation}
When $\lambda=0$ this equation gives the rate of change of the particle's energy as measured by a local observer, while the other three equations give the spatial components of the generalized Lorentz force. In any case, it is clear that the gravitational field has its influence because the geometry of spacetime (in this case, the metric) couples directly to the particle's 4-velocity as well as to the electromagnetic field. In principle, this equation can be used to test metric theories of gravity and for GW detection.

Consider for example the following simple application. For a given background metric we could compute the required experimental conditions, including the appropriate charge and current densities, such that the resulting electromagnetic fields are tuned to compensate the second term in Eq. (\ref{Lorentzforce}), i.e.,
\begin{equation}
F_{\mu\nu}g^{\mu\lambda}u^{\nu}=\dfrac{m}{q}\Gamma^{\lambda}_{\;\alpha\beta}u^{\alpha}u^{\beta} \quad  \Rightarrow  \quad  
F_{\mu\nu}g^{\mu\lambda}=\dfrac{m}{q}\Gamma^{\lambda}_{\;\alpha\nu}u^{\alpha}.\label{antigravity}
\end{equation}
In these conditions a particle would follow a straight path as if it were moving in a (pseudo) Euclidean spacetime. 
This local antigravity effect is testable, in principle, and is model dependent since the gravitational equations determine the background geometry. A more complete treatment would require the full Einstein-Maxwell equations (or similar systems), since if the Maxwell fields are strong enough to compensate gravity locally, then we should no longer disregard the back-reaction of these fields on spacetime geometry. Nevertheless, as an approximation we can still use the Maxwell equations for a specific fixed background geometry.
 
The fields can be tuned to the local geometry according to,
\begin{equation}
F_{\varepsilon\nu}=\dfrac{m}{q}g_{\varepsilon\lambda}\Gamma^{\lambda}_{\;\alpha\nu}u^{\alpha}.
\end{equation}
Inserting this expression in the inhomogeneous Maxwell equations, the corresponding charge-current sources generating the required fields are given by
\begin{eqnarray}
j^{\sigma}&=&\mu_{0}^{-1}\dfrac{m}{q}\Bigg[g^{\varepsilon\gamma}g^{\nu\sigma}\partial_{\gamma}(g_{\varepsilon\lambda}\Gamma^{\lambda}_{\;\alpha\nu}u^{\alpha})
+g_{\varepsilon\lambda}\Gamma^{\lambda}_{\;\alpha\nu}u^{\alpha} \times
    \nonumber  \\
&&\times \left[\partial_{\gamma}(g^{\varepsilon\gamma}g^{\nu\sigma})+\dfrac{1}{\sqrt{-g}}\partial_{\gamma}(\sqrt{-g})g^{\varepsilon\gamma}g^{\nu\sigma}\right]\Bigg].\label{fourcurrentantigrav} 
\end{eqnarray}

Considering, for simplicity, a diagonal metric, the electric and magnetic fields needed to balance gravity locally are 
\begin{eqnarray}
F_{j0}&=&-\dfrac{1}{c}E_{j}=\dfrac{m}{q}g_{jj}\Gamma^{j}_{\;\alpha 0}u^{\alpha},
\label{electricantigravity}
\end{eqnarray}
\begin{eqnarray}
F_{jk}&=&-B_{jk}=\dfrac{m}{q}g_{jj}\Gamma^{j}_{\;\alpha k}u^{\alpha},
 \label{electricmagneticantigrav}
\end{eqnarray}
respectively, where the only contraction is on $\alpha$. For example, suppose that in some reference frame a charge density distribution generates a single electric field. Assuming a diagonal metric, the required charge density might in principle be tuned to the local geometry according to
\begin{equation}
\rho=\dfrac{\varepsilon_{0}m}{q}u^{\alpha}\left[g^{kk}g^{00}\partial_{k}\left( g_{kk}\Gamma^{k}_{\;\alpha0}\right)-g_{kk}\Gamma^{k}_{\;\alpha0}\gamma^{k}\right].
\end{equation}
Here a sum over $k$ is assumed and $m$, $q$ and $u^{\alpha}$ are fixed parameters. This expression is compatible with the electric field expression in Eq. (\ref{electricantigravity}) for all $j$. Therefore, it gives the charge density required to maintain constant all $u^{j}$ components of the 3-velocity for the test object under the influence of both gravity and an electric field. If a magnetic field is present, such electric conditions are necessary but not sufficient according to the magnetic expressions in Eq. (\ref{electricmagneticantigrav}). 
In that case, we get
\begin{eqnarray}
j^{i}=-\dfrac{1}{\mu_{0}}\dfrac{m}{q}u^{\alpha}\left[ g^{ii}g^{jj}\partial_{j}\left(g_{ii}\Gamma^{i}_{\;\alpha j}\right)+g_{ii}\sigma^{ij}\Gamma^{i}_{\;\alpha j}\right].
\nonumber  \\
\label{jantigrav}
\end{eqnarray}
Notice that the expressions for the compensating fields  (\ref{electricmagneticantigrav}) were replaced inside Maxwell's inhomogeneous equations, consequently, we derived the necessary charge and current densities such that the anti-gravity could be achieved for a test body moving inside the charge and current distributions. Such applications could be implemented if the charge and current distributions were those of a charged fluid (as in the case of a plasma) in the presence of a gravitational field. If the conditions for the charge and current densities are satisfied locally at each point of the plasma where the test object is moving, then it will experience no gravity. This opens the possibility for a future technology in which the motion of the object could be controlled by controlling the local charge and current densities of the plasma surrounding the object. In particular, whenever the fields are higher than those expressed in Eq. (\ref{electricmagneticantigrav}) the object's motion will have an upward component.

\section{Summary and Discussion}\label{Conclusion}

\subsection{On the relation between electromagnetism and spacetime geometry in the foundations of electrodynamics}

The electromagnetic field equations based on the well-established postulates of charge and magnetic flux conservation laws and compatible with the Lorentz force are fully general and coordinate free, without requiring the metric or even the affine structure of the spacetime manifold \cite{Gronwald:2005tv,Hehl:2000pe,Hehl:2005hu,Hehl:1999bt}. What particularizes these equations for a given spacetime geometry are the constitutive relations between the field strenghts and the excitations. These relations in vacuum can be viewed as constitutive relations for the spacetime itself and necessarily introduce the conformal part of the metric \cite{Hehl:2005hu}. In fact, the causal structure of spacetime is intertwined with electrodynamics at the very foundational level.
If one modifies the constitutive relations new field equations, and consequently, new predictions for the electromagnetic phenomena follow. These relations are implicit in the action (or gauge) approach, therefore different relations imply different actions as in the cases of Heisenberg-Euler non-linear electrodynamics \citep{HeisenbergEuler} or Volterra-Mashhoon non-local electromagnetism \citep{Mashhoon}.
Assuming linear and local relations do not necessarily require homogeneity and isotropy. The electric permitivity and magnetic permeability tensors in vacuum are required inside the constitutive relations, in order to relate the physical dimensions of the field strengths and the excitations. Now, whereas in the laboratory the homogeneous and isotropic constitutive relations might seem to be valid (by measuring electric and magnetic fields through their effects on charges and testing the usual expressions for the inhomogeneous equations), it is not proven that such relations remain unchanged in the presence of strong gravitational fields. We suggest that the assumptions of homogeneity and isotropy might be inappropriate for physical situations in which the spacetime isometries transgress spatial homogeneity and/or isotropy.

 In the first section, we briefly addressed the issue that since these relations can be viewed as constitutive relations for the spacetime itself (or vacuum), these tensors can be interpreted to characterize the electromagnetic properties of spacetime (\cite{Hehl:2005hu,Lammerzahl:2004ww,Itin:2004qr,Hehl:2004yk}), or of what can be called the electro-vacuum.
 In this sense, two different, although related, issues deserve some debate. The first has a geometrical tone coming from the idea that if spacetime isometries are not {\it a priori} given, but must be considered locally for each astrophysical or cosmological scenario, then the same is expected for the symmetry properties of the permittivity and permeability tensors. The spacetime symmetries should be reflected in the components of these tensors, which in general, depend on the spacetime coordinates. This goes along with the line of reasoning of general relativity according to which, electromagnetic fields gravitate, affecting spacetime geometry, and propagate according to a law that depends on the local causal structure of spacetime. In this sense, without abandoning local conformal invariance, the {\it a priori} assumption of homogeneity and isotropy for the permittivity and permeability tensors can be abandoned. Consequently, according to these ideas, the velocity of light, determined by these electromagnetic properties of spacetime (or electrovacuum)  is predicted to be isotropic but inhomogeneous for spherically symmetric geometries (having a radial dependence), and inhomogeneous and anisotropic for axially symmetric cases (such as the cases of rotating relativistic stars or black holes). These predictions might be tested experimentally.
 
The second consideration that deserves a careful analysis is related to the idea that the physical properties of vacuum, spacetime geometry and electromagnetism seem to be deeply related as expressed in the constitutive relations. Operationally, these relations are required for the system of field equations to be complete and solvable, in principle. From the point of view of physical ontology these relations reinforce the idea of spacetime endowed with well-defined physical properties. Technically, the excitations are potentials for the charge and current distributions, as can be inferred using differential forms, and can be viewed as some sort of extended version of the so called sources, a fact that is clear from dimensional analysis. Therefore, physically the only way the fields can be causally linked to the charge and current distributions is via the constitutive relations which introduce the (conformal part of the) spacetime metric and the electromagnetic permittivity and permeability tensors. The link is achieved via physical spacetime and to introduce the notion of vacuum here is somehow unecessary if one accepts the idea that the spacetime manifold has a well-defined physical ontology. This notion is somehow reinforced according to the idea that spacetime is causally linked to mass-energy fields, becoming curved and affecting the field's propagation. On the other hand, according to the constitutive relations, strictly speaking it is the conformal part of spacetime geometry that might be said to have electromagnetic properties. 

One concludes that, in a very deep sense, the constitutive relations, are not only a technical detail of the electromagnetic theory (where the non-trivial cases are expected only for extraordinary non-linear or non-local effects, for example). These relations bring forward the debate on the very nature of space and time, of physical vacuum and its relations with electrodynamics.  

\subsection{On the physics of electromagnetism in curved spacetime}

To search observable effects of spacetime curvature (gravity) in electrodynamics, we assumed a particular form for the constitutive relations and considered the field equations in a general pseudo-Riemann background spacetime. For the homogeneous and isotropic relations (\ref{constitutive4d}) these field equations are simply affected via the metric components and its derivatives.
While the homogeneous equations are insensitive to the spacetime geometry of torsionless manifolds, the inhomogeneous equations are changed with many physical gauge invariant consequences.
In fact, electrostatics and magnetostatics are no longer separated, instead they become coupled due to the presence of curved geometry. Furthermore, new wave equations were derived.

For example, the coupling to spacetime geometry, in particular to the gravitomagnetic (time-space off-diagonal) terms, gives magnetic corrections to Gauss' law (\ref{GaussGenerall}). Therefore, the Gauss law in the background geometry of a rotating spherical mass necessarily includes magnetic terms, becoming electromagnetic.
 This coupling gives rise to the possibility of having even static magnetic fields as sources of electric fields. In astrophysical scenarios such as neutron stars with strong gravity, an induced electric field might arise due to the coupling between the magnetic field and the gravitomagnetism of the surrounding spacetime. This is an illustration of a gravitomagnetic effect affecting electromagnetism directly through the very nature of the field equations in curved spacetime. Therefore, there is an electromagnetic coupling via gravitomagnetic effects. The presence of magnetic fields in Gauss' law disappear for vanishing off-diagonal time-space metric components.

For a diagonal metric such couplings are no longer present, but for geometries where the functions $\gamma^{k}$ in (\ref{gausssimple}) are non-vanishing, the electric field in vacuum will necessarily be non-uniform, i.e., spacetime curvature will introduce an inevitable spatial variability in the electric field. In any case, dynamical (time varying) geometries, such as gravitational waves will affect electric fields via Gauss' law and induce electromagnetic waves. For example, a gravitational wave travelling in the $z$ direction with a (+) polarization corresponds to a diagonal metric and will affect an initially uniform electric field in the $z$ direction according to Gauss' law (\ref{gausssimple}). In vaccum, the electric field feels the dynamics of the travelling wave becoming time varying and non-uniform. Inside the charge distribution the electric field feels a similar effect but the gravitational wave will also affect the charge distribution according to Eq. (\ref{chargedynamical}).  

As in the case of Gauss' law, new predictions emerge in the Maxwell-Amp\`{e}re law due to the curvature of spacetime geometry (\ref{MaxAmpereGenerall}). For vanishing currents the presence of an electric field  can be a source of  magnetic fields, with an extra contribution to Maxwell's displacement current induced by spacetime dynamics, coming from the functions $\xi^{ii}$. These functions vanish for stationary spacetimes but might have an important contribution for strongly varying gravitational waves (high frequencies).
While in Gauss' law the electromagnetic coupling disappears for a diagonal metric, in the Maxwell-Amp\`{e}re law this coupling is always present. In particular, for dynamical (time varying) geometries, the two terms containing the electric field in (\ref{MaxAmpereGenerall}) come together since a non-stationary spacetime will necessarily induce a time varying electric field via the Gauss law. Accordingly, gravitational waves are expected to produce a direct effect in magnetic fields.

We also expect new phenomena associated with electromagnetic waves in the presence of a background gravitational field. It is natural to expect an effect in the polarization of electromagnetic waves induced by the curvature of (pseudo) Riemann spacetimes (see also \cite{Hacyan:2015kra}). For stationary geometries, the wave equations (\ref{wave stationary space}) and (\ref{wave magnetic stationary}) have no electromagnetic coupling. Nevertheless, the last three terms on the rhs of (\ref{wave stationary space}) and the second and fourth terms in the rhs of (\ref{wave magnetic stationary}) imply a coupling between the dynamics of the different field components. This fact suggests polarization effects induced by spacetime geometry.
The theory also seems to suggest longitudinal modes induced by curved spacetime geometry. These longitudinal modes in vector ($a^{k}$) wave equations appear whenever $\partial_{k}a^{k}\neq0$, and in fact, in usual electromagnetism such terms are absent because the fields in vacuum are divergenceless. In curved spacetime, we have $\partial_{k}E^{k}\neq0$ and this manifests in the terms containing the first derivatives of the electric field in Eq. (\ref{wave stationary space}). These longitudinal modes seem to be a prediction of electromagnetic wave dynamics in curved spacetime.   

In principle, the passage of a gravitational wave in a region with electromagnetic fields will have a measurable effect. To compute this we have to consider Maxwell's equations on the perturbed background of a gravitational wave. 
It is important to explore different routes towards gravitational waves detection and many alternatives might be explored beyond the Laser Interference techniques usually considered \citep{GWsRiles:2012yw}. Notice that the linearisation of gravity allows to derive the wave equations and that it is also the context in which gravitoelectric and gravitomagnetic fields can be defined \citep{GravitomagnetismMashhoon:2003ax}. In the Lorentz gauge, one can see that the ($\times$) polarization is responsible for effects due to (space-space) off-diagonal metric components, which resemble the gravitomagnetic effects that are present in metrics with time-space components. This brings interesting perspectives regarding the physical interpretations since the analogies with electromagnetism might be explored. In particular, gravitomagnetic effects on gyroscopes are known to be fully analogous to magnetic effects on dipoles. Now, in the case of gravitational waves these effects will in general be time-dependent. The tiny gravitomagnetic effect due to Earth's rotation on gyroscopes was successfully measured during the Gravity Probe B experiment \cite{Everitt:2011hp}, where the extremely small geodetic and Lens-Thirring (gravitomagnetic) deviations of the gyro's axis were measured with the help of Super Conducting Quantum Interference Devices (SQUIDS).  A possible application to GW detection could in principle result from similar effects, analogous to (time varying) gravitoelectromagnetic effects on gyroscopes, using SQUIDS. On the other hand, rotating superconducting matter seems to generate anomalous stronger gravitomagnetic fields (anomalous gravitomagnetic London moment) \cite{Tajmar:2004ww,Tajmar:2006gh} and if these results are robustly confirmed then superconductivity and superfluidity might somehow amplify gravitational phenomena. This speculation deserves further theoretical and experimental research as it could contribute for GW detection. 

Another promising route that we wish to emphasize in this work comes from the study of the coupling between electromagnetic fields and gravity. Are there measurable effects on electric and magnetic fields and related electromagnetic phenomena during the passage of a gravitational wave? Could these be used in practice to study the physics of gravitational wave production from astrophysical sources, or applied to gravitational wave detection? It seems reasonable to say that such a route is far from being fully explored. Regarding electromagnetic waves it has been shown that gravitational waves have an important effect on the polarization of light \cite{Hacyan:2015kra}. Are there other effects derivable from electromagnetic theory in curved space-time? We recall that lensing is gradually more and more relevant in observational astrophysics and cosmology and it seems relevant to study the effects on lensing due to the passage of gravitational waves from different types of sources. Could lensing provide a natural amplification of the gravitational perturbation signal due to the coupling between gravity and light? These topics need careful analysis for a better understanding of the possible routes (within the reach of present technology) for gravitational wave astronomy and its applications to astrophysics and cosmology.

In this work we pointed to some possible research lines regarding the effects of GW in electromagnetsm. In summary: GW induces electro-magnetic couplings in Gauss law and an extra term to Maxwell's displacement current in  Maxell-Amp\`{e}re law. Accordingly, GW induce eelectromagnetic waves with additional electro-magnetic couplings in the wave equations, with possible consequences for polarizations and interference patterns. Moreover, charge conservation implies that GWs induce electric charge oscillations. All these effects deserve further research due to its relevance for the physics of GW production and detection.  

\subsection{Applications and future work}

In a future work we will be addressing the applications of the generalized Gauss, Maxell-Amp\`{e}re and wave equations for different background geometries with astrophysical relevance, such as: spherically symmetric solutions, rotating solutions and gravitational waves. The following ideas are intended to establish a bridge to this second work under preparation \cite{FCFL}.

Electrodynamics in curved spacetime is expected to be important in certain relativistic astrophysics phenomena. Many astrophysical systems, from planets to stars and galaxies have magnetic fields which is somehow related to the fact that most astrophysical fluids are in the form of plasma. On the other hand, it seems that essentially all astrophysical bodies are rotating. In spite of this, the spherically symmetric solutions, such as the Schwarzschild metric of GR which we will consider, provide an important guideline for the physics of the gravitational-electromagnetic coupling. It is also a useful pedagogical tool to approach this topic. In some cases, the slow rotation regime can be applied, which brings an approximation to the Kerr metric giving the Schwarzschild solution plus an extra term depending linearly on the angular momentum. Such a term corresponds to the gravitomagnetic potential and can be responsible for some electromagnetic couplings explored in this work. 

It is well known that gravitomagnetism, in particular the Lens-Thirring frame dragging, seems to play an important role in the processes of collimation of astrophysical jets along well defined axis \citep{Jets,gravitomagneticBH, Chicone:2010hy}, in the so called Blandford-Znajek mechanism (\citep{kthorne,Penna:2014aza}). Such jets are detected in many active (radio) galaxies, revealing strong gravity and also strong magnetic fields due to the disk of very hot and quickly rotating plasma around the central black hole. The gravitational-electromagnetic coupling explored in the equations of electrodynamics in curved spacetime might be extremely important to deepen the understanding of such high energy processes. It should also be mentioned that the Lens-Thirring effect on electromagnetic fields, due to rotating supermassive black holes in the center of galaxies, might have another interesting application for the formation and evolution of galaxies. This effect can be understood as the result of a differential rotation of spacetime itself around the central object, analogous to the velocity field of hurricanes in planetary atmospheres, where the angular momentum is higher towards the central axis. Therefore, radial electric field lines in initial data will be distorted and are expected to produce a spiral pattern. To what extent such electromagnetic-gravitational effect can contribute for the understanding of the formation of spiral structures in galaxies remains an open question, motivating further research.

 We will be also considering the weak field limit using the Parameterized Post-Newtonian (PPN) formalism to make a bridge with the topic of testing alternative theories of gravity. Indeed, all the results in the present paper only assumed a (pseudo) Riemann spacetime geometry without any specification of the gravitational theory determining the metric functions. In principle, depending on the gravitational theory, the inhomogeneous equations and the wave equations will have different solutions and can therefore be used to constrain or invalidate different gravitational theories, which assume (pseudo) Riemann geometry. In the context of GR we will also consider GW as linear perturbations of Minkowski spacetime and explore physical, observable electromagnetic phenomena with potential relevance to deepen the evolving field of study of GW production and detection.

\section*{Acknowledgments}
F.S.N.L. acknowledges financial  support of the Funda\c{c}\~{a}o para a Ci\^{e}ncia e Tecnologia through an Investigador FCT Research contract, with reference IF/00859/2012, funded by FCT/MCTES (Portugal).

\end{document}